\pdfoutput=0
\documentclass[11pt]{article}
\usepackage{axodraw}
\usepackage{epsfig}
\usepackage{amsfonts}
\usepackage{amsmath}
\usepackage{bm,bbm}
\usepackage{cite}
\allowdisplaybreaks[1]

\input pix.sty

\newcommand{\la}[1]{\label{#1}}
\newcommand{\be}{\begin{equation}}
\newcommand{\ee}{\end{equation}}
\newcommand{\ba}{\begin{eqnarray}}
\newcommand{\ea}{\end{eqnarray}}
\newcommand{\rmi}[1]{{\mbox{\scriptsize #1}}}
\newcommand{\fig}{Fig.~}

\newcommand{\eq}{Eq.~}
\newcommand{\eqs}{Eqs.~}
\newcommand{\se}{Sec.~}
\newcommand{\ses}{Secs.~}
\newcommand{\nr}[1]{(\ref{#1})}

\newcommand{\nn}{\nonumber \\}
\newcommand{\fr}[2]{{\frac{#1}{#2}\,}}
\newcommand{\msbar}{{\overline{\mbox{\rm MS}}}}

\newcommand{\Lambdamsbar}{{\Lambda^{ }_{\mbox{\tiny\rm{IR}}}}}
\renewcommand{\vec}[1]{{\bf #1}}

\renewcommand{\eq}{eq.~}
\renewcommand{\eqs}{eqs.~}
\renewcommand{\se}{sec.~}
\renewcommand{\ses}{secs.~}
\renewcommand{\fig}{fig.~}

\newcommand{\PT}{\mathbbm{P}^\rmii{T}}

\newcommand{\Nf}{N_{\rm f}}
\newcommand{\Nc}{N_{\rm c}}

\newcommand{\gammaE}{\gamma_\rmii{E}}

\newcommand{\rmO}{{\mathcal{O}}}
\newcommand{\bmu}{\bar\mu}

\newcommand{\dA}{d_\rmii{A}^{ }}
\def\lsi{\raise0.3ex\hbox{$<$\kern-0.75em\raise-1.1ex\hbox{$\sim$}}}
\def\gsi{\raise0.3ex\hbox{$>$\kern-0.75em\raise-1.1ex\hbox{$\sim$}}}
\newcommand{\lsim}{\mathop{\lsi}}
\newcommand{\gsim}{\mathop{\gsi}}

\newcommand{\nB}{n_\rmii{B}}
\newcommand{\rmii}[1]{{\mbox{\tiny\rm{#1}}}}
\newcommand{\rmiii}[1]{{\mbox{\tiny{$\scriptstyle{\rm#1}$}}}}
\newcommand{\re}{\mathop{\mbox{Re}}}
\newcommand{\im}{\mathop{\mbox{Im}}}
\newcommand{\Tint}[1]{{\hbox{$\sum$}\!\!\!\!\!\!\!\int\,}_{\!\!\!\!\raise-0.9ex\hbox{$\scriptstyle{#1}$}}}
\newcommand{\Tinti}[1]{{{\Sigma}\!\!\!\!\raise0.3ex\hbox{$\int$}_\rmii{${#1}$}}}
\newcommand{\bi}{\begin{itemize}}
\newcommand{\ei}{\end{itemize}}
\newcommand{\hide}[1]{ }

\newcommand{\deltabar}{\raise-0.02em\hbox{$\bar{}$}\hspace*{-0.8mm}{\delta}}
\newcommand{\ddeltabar}{\raise-0.18em\hbox{$\bar{}$}\hspace*{-0.8mm}{\delta}}
\renewcommand{\P}{\mathcal{P}}
\renewcommand{\H}{\mathcal{H}}

\newcommand{\X}{\mathcal{X}}
\newcommand{\Y}{\mathcal{Y}}

\newcommand{\taui}{\tau_{i}} 
\newcommand{\taue}{\tau_{e}} 

\newcommand{\M}{\rmii{$M$}}
\newcommand{\T}{\rmii{$T$}}
\newcommand{\R}{\rmii{R}}
\newcommand{\mpl}{m_\rmii{pl}} 
\newcommand{\xiT}{\varrho_\rmii{$T$}} 
\newcommand{\xiQ}{\varrho_\rmii{$Q$}} 
\newcommand{\scalar}{h} 
\def\TAsc(#1,#2)(#3,#4,#5)%
{\SetWidth{2.0}\CArc(#1,#2)(#3,#4,#5)\SetWidth{1.0}}
\def\Lwidth{3}

\def\TAgl(#1,#2)(#3,#4,#5){\SetWidth{2.0}\PhotonArc(#1,#2)(#3,#4,#5){\Lwidth}%
{6.283 #3 mul 360 div #4 #5 sub #4 #5 sub mul sqrt mul Tdensity mul}%
\SetWidth{1.0}}
\def\TLgl(#1,#2)(#3,#4){\SetWidth{2.0}\Photon(#1,#2)(#3,#4){\Lwidth}
{#1 #3 sub #1 #3 sub mul #2 #4 sub #2 #4 sub mul add sqrt Tdensity mul}%
\SetWidth{1.0}}

\renewcommand{\picc}[1]{\;\parbox[c]{60pt}{\begin{picture}(60,30)(0,-10)
\SetWidth{1.0}\SetScale{1.3} #1 \end{picture}}\; }
\def\Lwidth{1.3}

\makeatletter \@addtoreset{equation}{section} \makeatother
\renewcommand{\theequation}{\arabic{section}.\arabic{equation}}
\makeatletter
\renewcommand\section{\@startsection {section}{1}{\z@}%
                                   {-5.5ex \@plus -1ex \@minus -.2ex}
                                   {2.3ex \@plus.2ex}%
                                   {\normalfont\large\bfseries}}
\renewcommand\subsection{\@startsection{subsection}{2}{\z@}%
                                     {-3.25ex\@plus -1ex \@minus -.2ex}%
                                     {1.5ex \@plus .2ex}%
                                     {\normalfont\normalsize\bfseries}}
\renewcommand\thesection {\@arabic\c@section}
\renewcommand\thesubsection   {\thesection.\@arabic\c@subsection}
\renewcommand{\@seccntformat}[1]{%
\csname the#1\endcsname.\hspace{1.0em}}
\makeatother

\begin{document}

\flushbottom

\begin{titlepage}

\begin{flushright}
November 2022
\end{flushright}
\begin{centering}
\vfill

{\Large{\bf
    Gravitational wave background from vacuum and thermal  \\[3mm]
    fluctuations
    during axion-like inflation
}} 

\vspace{0.8cm}

P.~Klose, 
M.~Laine, 
S.~Procacci

\vspace{0.8cm}

{\em
AEC, 
Institute for Theoretical Physics, 
University of Bern, \\ 
Sidlerstrasse 5, CH-3012 Bern, Switzerland \\}

\vspace*{0.8cm}

\mbox{\bf Abstract}
 
\end{centering}

\vspace*{0.3cm}
 
\noindent
We revisit the framework of axion-like
inflation in view of the possibility
that the coupling of the inflaton to a non-Abelian topological charge
density could lead to the generation of a rapidly thermalizing heat bath.
Both dispersive (mass) and absorptive (friction) 
effects are included. For phenomenologically 
viable parameters, the system remains in a weak 
regime of warm inflation (thermal friction $\ll$ Hubble rate). 
For tensor perturbations we derive an interpolating formula  
that incorporates both vacuum and thermal production. 
The latter yields a model-independent frequency shape $\sim f_0^3$ 
in the LISA window, whose coefficient allows to measure
the maximal shear viscosity of the thermal epoch.  It is a challenge,
however, to find models where the coefficient is 
large enough to be observable. 

\vfill


\end{titlepage}

\tableofcontents

%
\section{Introduction}
\la{se:intro}

In order to make maximal use of the upcoming 
LISA gravitational wave interferometer,  
it is important to have frequency templates for various
phenomena that might take place in the early universe.
Among the most important sources of gravitational waves
is inflation~\cite{gw_infl}. 
The corresponding energy density 
is constrained by the indirect effect of the 
gravitational background on CMB photons, parametrized 
by the tensor-to-scalar ratio $r$, which is bounded from above
at small frequencies 
($f^{ }_0 \ll 10^{-15}_{ }$~Hz today\hspace*{0.2mm}\footnote{%
 The scales probed by the CMB are normally expressed via
 the wave number, 
 $p^{ }_0 \in (10^{-4}_{ } - 10^{-1}_{ })$~Mpc$^{-1}_{ }$~\cite{planck}. 
 \la{Mpc}
 }). 
It is possible that physics at a late stage of inflation
leads to an additional 
contribution~\cite{gw_preheat1,gw_preheat2,gw_preheat3,gw_preheat4}, 
which could increase with frequency. 
The LISA sensitivity peaks 
at $f^{ }_0 \sim (10^{-4}_{ } - 10^{-1}_{ })$~Hz~\cite{gn}, 
and ideally we could envisage a mechanism that 
sources an observable spectrum in this domain. 
 
One example of such a mechanism is 
axion inflation~\cite{ai,ai_em,ai_rev}. 
In this case, it has been argued that 
an efficient tachyonic instability
could convert a significant fraction of the initial 
energy density into gravitational 
radiation~\cite{axion_gw_1,axion_gw_2,axion_gw_3}
(cf.,\ e.g.,\ ref.~\cite{axion_gw_4} for an overview). 
Therefore this is one of the models for which 
LISA sensitivity studies have
been carried out~\cite{axion_lisa}.  

It could be asked, however, how robust this scenario is. 
Notably, many studies concern the case of an 
axion coupling to Abelian gauge fields. The large
signal originates when their conversion into 
gravitational waves proceeds
without any backreaction. But backreaction 
must exist, even just from energy conservation
(cf.,\ e.g.,\ refs.~\cite{axion_br_energy,axion_br_simu} 
and references therein), 
and quite concretely if the Abelian fields
are coupled to fermions~\cite{axion_br_fermions}. 

If the gauge sector is non-Abelian, which appears 
natural from the particle physics point of view, backreaction
is more substantial, as it takes
place even without fermions.  
Such a system has been argued to interact strongly enough to 
thermalize already during inflation~\cite{fixed_pt}.\footnote{%
 The obstacles for warm inflation raised in ref.~\cite{linde} 
 are evaded in the case of non-Abelian axion inflation, 
 as reviewed in the introduction of ref.~\cite{warm}. 
 }
Thermalization maximizes backreaction,
by eliminating memory of initial conditions. 
In this case, the fraction of energy density converted
to gravitational radiation is not of order unity, 
but it is rather suppressed by couplings~\cite{gravity}. 
It appears interesting to 
estimate whether an observable signal could originate nevertheless. 

Our goal is to study gravitational wave production 
in the case of a thermalized non-Abelian gauge sector. 
Previously, we have done this for the signal originating
at reheating~\cite{gravity}, 
which peaks at high frequencies, 
$f^{ }_0 \sim 10^{11}_{ }$~Hz, and can
be constrained by the observable $N^{ }_\rmi{eff}$.\footnote{%
 This parametrizes the overall energy density throuh the 
 effective number of relativistic ``neutrino-like'' species, 
 so that a non-zero $N^{ }_\rmii{eff} - 3 $ originates either from 
 non-trivial Standard Model dynamics, or from 
 beyond the Standard Model degrees of freedom. \la{fn:Neff}
 } 
Now we turn to gravitational wave 
production during the inflationary stage itself, 
with particular focus on 
the LISA frequency window, $f^{ }_0 \sim (10^{-4}_{ } - 10^{-1}_{ })$~Hz.

\vspace*{3mm}

To be specific, 
we consider a theory defined in local Minkowskian coordinates
by the Lagrangian
\be
 \mathcal{L} 
 \; \supset \;
 \frac{ 1 }{2} \partial^\mu\varphi\, \partial_\mu\varphi
 - V^{ }_0(\varphi)  
 - \frac{\varphi \,  \chi}{f^{ }_a}
 \;, 
 \quad
 \chi \;\equiv\;
 c^{ }_\chi \, 
 \epsilon^{\mu\nu\rho\sigma}_{ }
  g^2 F^{c}_{\mu\nu} F^{c}_{\rho\sigma}
 \;, 
 \quad
 c^{ }_\chi \; \equiv \; 
 \frac{1}{64\pi^2}
 \;, \la{L}
\ee
where $\varphi$ is the inflaton field, 
$
 F^c_{\mu\nu} \equiv \partial^{ }_\mu A^c_\nu - \partial^{ }_\nu A^c_\mu
 + g f^{cde}_{ }A^d_\mu A^e_\nu
$ 
is the Yang-Mills field strength, 
$\Nc^{ }$ is the number of colours, 
$ c \in \{ 1,..., \dA \}$,
$ \dA \; \equiv \; \Nc^2 - 1 $, 
$g^2 \equiv 4\pi\alpha$,   
and $f^{ }_a$ is the axion decay constant. 
We stress that our $\varphi$ 
is not a physical QCD or dark matter
axion, but a more general field, 
heavier than would-be dark matter candidates, and unstable. 

The potential $V^{ }_0(\varphi)$ appearing 
in \eq\nr{L} requires some elaboration. 
We consider $V^{ }_0(\varphi)$
to be a ``bare'' vacuum potential, whereas  
$V(\varphi)$ denotes a potential including dynamical effects
from gauge field interactions
(including thermal effects if the system has thermalized).
Omitting thermal effects for the moment
(these are added in \se\ref{ss:reCR})
and postponing variations of the potential till later on 
(cf.\ \se\ref{ss:other}), 
we carry out our baseline discussion with the ansatz
\be
 V(\varphi)
 \;\simeq\;
 m^2 f_b^2\, \biggl[ 1 - \cos\biggl( \frac{\varphi}{f^{ }_b} \biggr) \biggr]
 \;. \la{V}
\ee
In \eq\nr{V}, 
$\sqrt{ m f^{ }_b } \sim \Lambda^{ }_\rmiii{UV}$ 
corresponds to the confinement
scale of some unified theory. Technically, we assume that 
inflationary physics has an energy scale 
$\epsilon^{ }_\varphi \ll \Lambda^{ }_\rmiii{UV}$, 
but that $\epsilon^{ }_\varphi$ is much 
larger than the confinement scale $\Lambda^{ }_\rmiii{IR}$
of some unbroken SU($\Nc^{ }$) subgroup, 
whose coupling $\alpha\ll 1$ is weak. It is the latter gauge group
that has been displayed in \eq\nr{L}.

On the semiclassical level, the periodicity of \eq\nr{V} is related
to the topological character of $\chi$ in \eq\nr{L}, i.e.\ that
$\chi$ evaluates to an integer for smooth vacuum configurations. 
This implies that 
$f^{ }_b = f^{ }_a$, and we assume this relation throughout. 
Given the operator in \eq\nr{L}, 
radiative corrections to $V(\varphi)$
are suppressed by powers of $1/f^{2}_a$, 
and appear on par with effects from 
higher-dimensional operators that have been  
omitted from \eq\nr{L}. 
However, infrared (IR) contributions, 
such as thermal corrections, 
can be consistently included at $\rmO(1/f^2_{a})$. 

\vspace*{3mm}

This paper is organized as follows. 
We start by considering the semiclassical evolution of the 
average inflaton field,
including thermal friction and mass corrections
(cf.\ \se\ref{se:background}). Then we discuss how this 
background sources tensor perturbations, or gravitational waves 
(cf.\ \se\ref{se:tensor}). After analyzing the shape of 
the corresponding spectrum (cf.\ \se\ref{se:numerics}), 
conclusions can be offered (cf.\ \se\ref{se:concl}).
Appendix~A elaborates on ``dispersive'' 
relations between thermal friction and mass effects, 
implying that both should be considered on equal footing. 
  
%
\section{Background solution}
\la{se:background}

Assuming that the gauge fields appearing 
in \eq\nr{L} thermalize rapidly, they form a radiation bath to which
the inflaton couples. As a first step, we recall the resulting 
evolution equations, yielding what is called the background solution.  
This topic agrees technically with 
the setup of warm inflation~\cite{setup_warm_1,setup_warm_2}, 
recently studied with renewed interest in the axion inflation 
context~\cite{warm_axion1,warm_axion2,warm_axion3,warm,warm45,warm_axion5,%
warm_axion6}.
However, thermal mass corrections (cf.\ \se\ref{ss:reCR}) have often 
been omitted, so we devote particular attention to this aspect.

Before proceeding, it is important to note that the 
corrections originating from the gauge field heat bath depend not 
only on the temperature, but also on the curvature of the space in
which the gauge fields live, which is parametrized by the 
Hubble rate~$H \equiv \dot{a}/a$, where $a$ is the scale factor
in Friedmann-Robertson-Walker coordinates. 
Thermal effects can be consistently included if 
the thermal scattering or equilibration rate, $\sim \alpha^2 T$, 
is large compared with the Hubble rate, $\alpha^2 T \gsim H$. 
In this regime, 
corrections proportional to the Hubble rate are small compared
with the thermal ones, $H^2 \ll T^2$. The computations are carried out
in a local Minkowskian frame, and subsequently  
the equivalence principle is invoked in order to  
convert the results to a general frame. If the dynamical solution leads to 
a smaller temperature, then the assumption of a thermal
medium is not self-consistent. 
But in this regime 
thermal corrections are very small, 
so their incomplete inclusion entails only 
an insignificant error. 

%
\subsection{Equations of motion}
\la{ss:eom}

To get the background solution, 
we consider a homogeneous average field, 
$\varphi(t,\vec{x})\to \bar{\varphi}(t)$, 
and couple the inflaton to the energy-momentum tensor of a plasma.
Simultaneously we include thermal corrections 
(whose form is specified in \se\ref{ss:reCR}) in the effective potential
from \eq\nr{V}. Then the classical equations of motion reduce to 
\ba
 \ddot{\bar\varphi} + (3 H + \Upsilon)\dot{\bar\varphi} +  V^{ }_\varphi 
 & \simeq & 
  0
 \;, \la{eom_field} \\[2mm] 
 \dot{e}^{ }_r + 3 H \bigl( e^{ }_r + p^{ }_r - T  V^{ }_\T \bigr)
 - T \dot{V}^{ }_\T  
 & \simeq &
 \Upsilon^{ }
 \dot{\bar\varphi}^2
 \;, \la{eom_plasma}
\ea
where $V^{ }_x \equiv \partial^{ }_x V$.
The form of these equation ensures overall energy conservation, 
with a gradual transfer of 
energy from $\bar\varphi$ to radiation.\footnote{%
 The opposite process, stochastic transfer of energy from the plasma
 to $\varphi$, can be included through a noise term in 
 \eq\nr{eom_field}~\cite{hydro}. This is important for the origin 
 of scalar perturbations, but plays no practical role for the 
 background solution, or for the tensor perturbations 
 discussed in \se\ref{se:tensor}. 
 }
The Friedmann equation closes the system, 
\be
 H \; = \; \frac{\dot{a}}{a} \; = \; \sqrt{\frac{8\pi}{3}} 
 \frac{\sqrt{ e^{ }_r + V -T  V^{ }_\T  
  + \dot{\bar\varphi}^2/2 }}
      { \mpl^{ } }
 \;, \la{eom_friedmann} 
\ee
where we have parametrized the Newton constant as $G \equiv 1/\mpl^2$, 
$\mpl^{ } = 1.22091 \times 10^{19}_{ }$~GeV.

As for $e^{ }_r$ and $p^{ }_r$, we note that at temperatures 
below the confinement scale of the SU($\Nc^{ }$) group
in \eq\nr{L} ($T\ll \Lambda^{ }_\rmiii{IR}$), 
the heat capacity is exponentially
small. Then, a small release of entropy leads to a large change in
the temperature. Therefore, we may 
assume that the system has already heated
up to a temperature above $ \Lambda^{ }_\rmiii{IR} $
(which should be minuscule compared with the Planck scale, 
or the ultraviolet
confinement scale $\Lambda^{ }_\rmiii{UV}$ mentioned below \eq\nr{V}). 
This simplifies
matters, as we may use a conformal equation of state for radiation,
$p^{ }_r \simeq g^{ }_* \pi^2 T^4/90$. 
For $T\gg \Lambda^{ }_\rmiii{IR}$,
we can assume $\alpha\ll 1$, 
which is beneficial for 
our analysis (cf.\ the discussion below \eq\nr{c}). 

Apart from the evolution of the inflaton and the temperature, 
dictated by \eqs\nr{eom_field} and \nr{eom_plasma}, we also solve
for the increase of the number of $e$-folds, 
$
 N \equiv \ln (a / a^{ }_\rmi{ref})
$, 
from 
\be
 \dot{N} = H \;. \la{eom_N}
\ee
Here $a^{ }_\rmi{ref}$ denotes the scale parameter at some
initial time, $t = t^{ }_\rmi{ref}$, 
specified for the numerics later on
(cf.\ caption of \fig\ref{fig:benchmark}).
Furthermore it is important to know how the ratio of a co-moving
momentum to the Hubble rate, $k/(aH)$, changes with time. 
This can be expressed as 
\be
 \frac{k}{(aH)(t)} = 
 \frac{k\, e^{-N(t)}_{ } }{a^{ }_\rmi{ref}\, H(t)} 
 \;. \la{eom_koaH} 
\ee
As long as $H\approx$ constant, this ratio decreases exponentially. 
Once inflation ends, $H$ starts to decrease and $N$ is almost constant,
whence the ratio increases (cf.\ \fig\ref{fig:benchmark}). 
The factor $a^{ }_\rmi{ref}$ drops out later on, when physical results are 
expressed in terms of present-day observables.

As a technical remark, 
we note that if we integrate over a long period of time,
interpolating between the inflationary and the radiation epochs, 
then it is favourable to replace $t$ by another variable
in \eqs\nr{eom_field} and \nr{eom_plasma}. 
For this purpose,  we employ in \se\ref{se:numerics}
\be
 x - x^{ }_\rmi{ref} \;\equiv\; 
 \ln\biggl( \frac{ t }{ t^{ }_\rmi{ref} } \biggr)
 \;. \la{def_x}
\ee

%
\subsection{Thermal friction coefficient}
\la{ss:imCR}

A key role in \eqs\nr{eom_field} and \nr{eom_plasma} is played
by the friction coefficient, $\Upsilon$. 
It needs to be fixed as it originates from 
the operator in \eq\nr{L}, which couples
the inflaton to the heat bath.
The value is determined by the response function 
of the medium evaluated at the appropriate frequency scale~\cite{warm}, 
thereby incorporating both vacuum physics 
(for $\omega \sim m \gg \pi T$) and thermal physics. 
The latter is important particularly at high temperatures, 
if we are driven 
to the regime $\omega \,\lsim\, m \,\lsim\, \alpha\Nc^{ } T$. 

It was indicated above that the frequency scale $\omega$ was chosen
to coincide with the mass scale at the minimum of the potential, $m$. 
As an alternative, we have also tested a dynamically evolving $\omega$, 
defined as 
$
 \omega_\rmi{dyn}^{ } \; \equiv \; \sqrt{\max(0,V^{ }_{\varphi\varphi})}
$. 
Even if this does have an effect, influencing the results quantitatively
on the $\rmO(1)$ level, we have not found any qualitative changes, 
and therefore stick to the simpler choice
$\omega = m$ in the following. 

It has been appreciated in 
refs.~\cite{warm_axion1,warm_axion2,warm_axion3,warm,%
warm45,warm_axion5,warm_axion6} 
that determining the value of $\Upsilon$
in the regime $\omega \,\lsim\, \alpha\Nc^{ } T$
requires non-perturbative input. Indeed 
the value at $\omega\to 0$ can be taken over 
from older lattice simulations~\cite{mt}, while the broader shape, extending
to $\omega \sim \alpha\Nc^{ } T$, has only been 
measured recently~\cite{clgt}. Larger frequencies can 
be addressed with a perturbative computation~\cite{Bulk_wdep}, 
and subsequently 
the parts can be put together,
\ba
 \Upsilon  & \simeq &  
 \frac{\dA \alpha^2}{f_a^2}
 \biggl\{  
 \kappa\, ( \alpha \Nc^{ } T)^3_{ } \, 
 \frac{1 + 
   \frac{\omega^2}
        { (  c^{ }_\rmiii{IR} \alpha^2 \Nc^2 T )^2_{ } }
 }{ 1 + 
   \frac{\omega^2}{ (  c^{ }_\rmiii{M} \alpha \Nc^{ } T )^2_{ } }
 }
 + 
 \Bigl[ 1 + 2 \nB^{ }\Bigl( \frac{\omega}{2} \Bigr)\Bigr]
 \frac{ \pi\omega^3 }{(4\pi)^4} 
 \biggr\} 
 \;, \la{Upsilon_interpolation} \\[2mm]
 \kappa           \!\! & \simeq & \!\! 1.5 \;, \quad
 c^{ }_\rmiii{IR} \; \simeq \; 106 \;, \quad
 c^{ }_\rmiii{M}  \; \simeq \; 5.1 
 \;, 
\ea
where $\nB^{ }(x) \equiv 1 / (e^{x/T} - 1)$ is the Bose distribution. 
We fix $\Nc^{ } = 3$ for numerical
illustrations and approximate the coupling according to \eq\nr{gg}.

%
\subsection{Thermal mass correction}
\la{ss:reCR}

The potential $V$ in \eqs\nr{eom_field}--\nr{eom_friedmann} 
can be partly parametrized by its curvature around the global minimum, which
we call the mass squared. The full shape contains other parameters as well, 
and their treatment is not as easy to handle in a universal manner; 
further comments on this follow below \eq\nr{c}. 
The mass squared experiences thermal corrections, 
which can formally be related to the real part of the same 
retarded correlator, $C^{ }_\R$, 
whose imaginary part determines $\Upsilon$~\cite{warm}.
In appendix~A 
we summarize general issues associated with 
the relation of the real and imaginary parts of $C^{ }_\R$, 
whereas in the remainder of the present section, 
we show how $\re C^{ }_\R$ can be determined 
through a perturbative computation.

A compact expression for $ C^{ }_\R $ 
can be given within the imaginary-time formalism
of thermal field theory. From \eq(3.2) of ref.~\cite{Bulk_ope}, 
after inverting the overall sign to conform with Minkowskian conventions, 
we have
\be
 C^{ }_\R(\omega + i 0^+_{ })
 = 
 8 \dA c_\chi^2 (D-3)(D-2) g^4 
 \Tint{Q} 
 \biggl\{ 
  \frac{K^4}{Q^2(Q-K)^2} - \frac{2 K^2}{Q^2}
 \biggr\}^{ }_{K\to (-i[\omega + i 0^+_{ }],\vec{0})} 
 \; + \; 
 \rmO(g^6) 
 \;, 
\ee
where $D \equiv 4 - 2\epsilon$ is the dimension of spacetime 
and $\Tinti{Q}$ denotes a bosonic Matsubara sum-integral. After 
carrying out the Matsubara sum, the result contains terms with 
and without the Bose distribution, $\nB^{ }$. The terms without $\nB^{ }$
are vacuum contributions, and can be determined analytically, 
leading to an UV divergent result, 
which requires renormalization. 
The terms with Bose distributions
are finite, 
but need to be evaluated numerically. 

We obtain, 
with 
$
  \bmu^2 \equiv 4\pi \mu^2 e^{ - \gammaE }_{ }
$
denoting the $\msbar$
renormalization scale, 
\be
  \re C^{ }_\R (\omega ) = 
  \frac{\dA c_\chi^2 g^4 \omega^2}{\pi^2 f_a^2}
 \biggl\{ 
   \omega^2 \mu^{-2\epsilon}_{ }
   \biggl(
     \frac{1}{\epsilon} + \ln\frac{\bmu^2}{\omega^2} - 1 
   \biggr)
   + 
   \int_0^\infty \! {\rm d}q \, \mathbbm{P} \, \biggl[ 
   \frac{16 q^3 \nB^{ }(q)}{(q-\frac{\omega}{2})(q + \frac{\omega}{2})} 
   \biggr]
 \biggr\} 
 \;, \la{reCR}
\ee
where $\mathbbm{P}$ denotes the principal value. 
The vacuum part agrees with 
\eq(5.2) of ref.~\cite{warm}. 
The thermal part 
has the limiting values
$
 \re C^{ }_\R (\omega ) |^{ }_\T 
  \stackrel{ T \;\ll\; \omega   }{\to}
 - {64} \dA c_\chi^2 g^4 \pi^2 T^4 / ( {15 f_a^2} )
$
and
$
 \re C^{ }_\R (\omega ) |^{ }_\T 
  \stackrel{ T \;\gg\; \omega }{\to}
  {8} \dA c_\chi^2 g^4 \omega^2 T^2 / ( {3 f_a^2} )
$, and 
it crosses zero at $\omega \approx 5.2 T$.

The effective mass squared is given by~\cite{warm}
\be
 m_\T^2 \;\approx\; m^2 - \re C^{ }_\R(m)
 \;. \la{mmT} 
\ee
The effective theory description is self-consistent provided that
$\re C^{ }_\R(m) \ll m^2$. While this is certainly true for
$T \ll m$, the constraint becomes non-trivial for $T \gg m$, 
where $\re C^{ }_\R(m) > 0$. Inserting the limiting
value from below \eq\nr{reCR}, $m_\T^2$ remains positive for 
$\alpha^2 T^2 \ll f_a^2$, which is just the condition
for the validity of the effective theory.

For \eqs\nr{eom_plasma} and \nr{eom_friedmann}, we need the first
and second temperature derivatives of the effective mass squared, 
\ba
 \partial^{ }_\T m_\T^2 & = & 
 - \frac{16 \dA c_\chi^2 g^4 m^2}{\pi^2 f_a^2 T^2}
   \int_0^\infty \! {\rm d}q \, \mathbbm{P} \, \biggl\{ 
   \frac{q^4 \nB^{ }(q) [ 1 + \nB^{ }(q) ]}
   {(q-\frac{m}{2})(q + \frac{m}{2})} 
   \biggr\}
  \;, \la{dTmmT} 
  \\ 
 \partial^{2}_\T m_\T^2 & = & 
 - \frac{16 \dA c_\chi^2 g^4 m^2}{\pi^2 f_a^2 T^3}
   \int_0^\infty \! {\rm d}q \, \mathbbm{P} \, \biggl\{ 
   \frac{q^5 \nB^{ }(q) [ 1 + \nB^{ }(q) ]}
   {(q-\frac{m}{2})(q + \frac{m}{2})} 
   \biggl[ 
     \frac{1+2\nB^{ }(q)}{T} - \frac{2}{q}
   \biggr]
   \, \biggr\} 
  \;. \la{ddTmmT}
\ea
Given the asymptotic values mentioned below \eq\nr{reCR}, 
both are positive for $T \ll m$ and 
negative for $T \gg m$; they again cross zero at $m\sim2\pi T$.
The physical quantities in which \eqs\nr{dTmmT} and \nr{ddTmmT} appear, 
are the entropy density and the heat capacity, respectively, 
\ba
 \bar{s} & = & s^{ }_r - V^{ }_\T
 \; \simeq \;
 \frac{ 2 g^{ }_{*} \pi^2_{ } T^3_{ } }{ 45 } - \partial^{ }_\T m_\T^2\, 
 f_b^2 \biggl[ 1 - \cos\biggl( \frac{ \bar\varphi }{ f^{ }_b } \biggr) \biggr]
 \;, \la{s} \\ 
 \bar{c} & = & c^{ }_r - T V^{ }_{\T\T} 
 \; \simeq \;  
 \frac{ 2 g^{ }_{*} \pi^2_{ } T^3_{ } }{ 15 } - T \partial^{2}_\T m_\T^2\, 
 f_b^2 \biggl[ 1 - \cos\biggl( \frac{ \bar\varphi }{ f^{ }_b } \biggr) \biggr]
 \;. \la{c}
\ea

For the second steps in \eqs\nr{s} and \nr{c}, 
we have assumed that the functional form of \eq\nr{V} remains
intact in the thermal corrections, i.e.\ that the parameter
$f_b^2$ does not get corrected. The logic here is 
that its inverse $1/f_b^2$ represents 
the expansion parameter of the effective theory treatment, 
and that corrections to it would only arise at
order $1/f_b^4$. However, other philosophies could be 
envisaged, for instance that we subtract from $V$ 
the vacuum quadratic part and only apply the thermal
correction to this one, 
\be
 V^\rmi{alt}_{ } \; \simeq \; 
 m^2 f_b^2\, 
 \biggl[ 1 - \cos\biggl( \frac{\bar\varphi}{f^{ }_b} \biggr) \biggr]
 + \frac{m_\T^2 - m^2}{2}\, \bar\varphi^2 
 \;. \la{V_alt}
\ee
The corresponding 
$
 \bar{s}
$
and
$
 \bar{c}
$
are like in \eqs\nr{s} and \nr{c}, 
but with 
$
 1 - \cos(\bar\varphi / f^{ }_b)
$
expanded to quadratic order in $\bar\varphi$. 
We have checked that our numerical results in 
\se\ref{se:numerics} remain unaltered if we replace
$V$ by $V^\rmi{alt}_{ }$.

For a consistent treatment, 
both \eq\nr{s} and \nr{c} should remain positive. 
The tricky domain is that of low temperatures, where 
\eqs\nr{dTmmT} and \nr{ddTmmT} are positive. 
Recalling the asymptotics
from below \eq\nr{reCR} and the minus sign from \eq\nr{mmT}, 
and setting $f^{ }_a = f^{ }_b$, we see that the relative 
corrections are small provided that $\alpha^2 \ll 1$. 
This assumption we are making in any case, as we are employing
a non-interacting form for the radiation contributions $s^{ }_r$ and $c^{ }_r$
to $\bar{s}$ and $\bar{c}$, respectively
(cf.\ the discussion in \se\ref{ss:eom}). 

For the practical computations, we need to insert a value for the 
gauge coupling. Following ref.~\cite{warm}, \eq\nr{reCR} is renormalized 
in the $\msbar$ scheme, setting $\bmu\to f^{ }_a ( = f^{ }_b)$, whereas 
the coupling is approximated as 
\be
 g^2 \;\simeq\; \frac{3(4\pi)^2}{22\Nc^{ }}
 \ln^{-1}_{ }\biggl[ \frac{\sqrt{(2\pi\Lambdamsbar)^2 + (2\pi T)^2 + m^2}}
                   {\Lambdamsbar}
             \biggr]
 \;. \la{gg}
\ee
The initial temperature is taken from the domain 
$T \gg \Lambdamsbar$, guaranteeing that $\alpha \ll 1$
(we have set $\Lambdamsbar \equiv 0.2$~GeV). 
We remark that the effective $T$-dependence of $g^2$ is
formally a higher-order effect, of $\rmO(g^4)$, 
and therefore omitted in \eqs\nr{dTmmT} and \nr{ddTmmT}. 

%
\section{Tensor perturbations}
\la{se:tensor}

Having determined the background solution in 
\se\ref{se:background}, we now turn to perturbations. 
While the treatment of scalar perturbations is well established, 
that of tensor perturbations is slightly less so, 
so we go through a detailed exposition. 

%
\subsection{Vacuum fluctuations in de Sitter spacetime}
\la{ss:vacuum}

We start by reviewing the fluctuation spectrum of
a massless\hspace*{0.2mm}\footnote{%
  We are implicitly assuming that there is a (gauge) symmetry which
  protects the field from receiving mass corrections, as this 
  is the case that can be applied to tensor perturbations.
 } 
scalar field in de Sitter spacetime,  
as it turns out that tensor perturbations can be taken over
from this result with minor modifications
(cf.\ \se\ref{sss:helicity}). 
First the standard derivation 
is repeated in a minimal manner (cf.\ \se\ref{sss:canonical}),  
and subsequently we carry it out 
with the so-called stochastic formalism~\cite{stochastic}
(cf.\ \se\ref{sss:stochastic}), 
as this helps to incorporate 
thermal fluctuations 
(cf.\ \ses\ref{ss:thermal} and \ref{ss:both}). 

%
\subsubsection{Canonical derivation}
\la{sss:canonical}

Considering a massless scalar field $\scalar$, we first need to 
fix its normalization. Canonical normalization is specified by giving
the action or the Hamiltonian in a local Minkowskian frame, 
\be
 \mathcal{S}^{ }_\M = 
 - \int^{ }_\Y \frac{1}{2}\, 
 \scalar^{,\mu}_{ }\, \scalar^{ }_{,\mu}
 \;, \quad
 {H}^{ }_\M = 
 \int \! {\rm d}^3\vec{y} \, 
 \frac{1}{2} 
 \bigl[ \, 
 \dot{\scalar}^2 + (\nabla\scalar)^2 
 \, \bigr] 
 \;, \la{normalization}
\ee 
where $\Y \equiv (t,\vec{y})$, 
and we have adopted the ``mostly plus''
metric convention. 
The quantized on-shell field operator is 
\be
 \bigl[\, \scalar \,\bigr]^{ }_\M
 = 
 \int \! \frac{{\rm d}^3\vec{p}}{\sqrt{(2\pi)^{3}_{ } 2 p}} \, 
 \Bigl[ \,
    a^{ }_\rmii{\vec{p}}\, e^{-i p t + i \vec{p}\cdot\vec{y}}_{}   
  + 
    a^{\dagger}_\rmii{\vec{p}}\, e^{i p t - i \vec{p}\cdot\vec{y}}_{}   
 \, \Bigr]^{ }_\M
 \;, \la{varphi_quantized_M}
\ee
where the annihilation and creation operators 
$a^{ }_\rmii{\vec{p}}$ and $a^{\dagger}_\rmii{\vec{p}}$ 
satisfy the commutator in \eq\nr{commutator}, and
$[\,]^{ }_\M$ is a reminder of the momenta being Minkowskian. 

For future reference, 
let us express the field operator in \eq\nr{varphi_quantized_M} in 
co-moving conformal coordinates. We denote co-moving momenta 
by $\vec{k}$ and $\vec{l}$, their relations to physical
momenta being $\vec{p} = \vec{k} / a$ and $\vec{q} = \vec{l} / a$. 
The co-moving coordinate $\vec{x}$ corresponds to the physical 
$\vec{y} = a \vec{x}$, whereas the conformal and physical times 
are related by 
$
 {\rm d}\tau = {\rm d}t / a 
$.
The canonical commutator becomes
\be
 [\, a^{ }_\rmii{\vec{p}} , a^{\dagger}_\rmii{\vec{q}} \,]
 \; = \; 
 \delta^{(3)}(\vec{p-q})
 \; = \; 
 a^3 \, \delta^{(3)}(\vec{k-l})
 \; \equiv \; 
 a^3 \, 
 [\, w^{ }_\rmii{\vec{k}} , w^{\dagger}_\rmii{\vec{l}} \,]
 \;. \la{commutator}
\ee
Implementing the coordinate transformation 
yields what will later on be  
interpreted as the field operator in the distant past, namely 
\be
 \bigl[\, \scalar \,\bigr]^{ }_\M
 \; = \; 
 \frac{1}{a}
 \int \! \frac{{\rm d}^3\vec{k}}{\sqrt{(2\pi)^{3}_{ } 2 k}} \, 
 \Bigl[ \,
    w^{ }_\rmii{\vec{k}}\, e^{-i k \tau + i \vec{k}\cdot\vec{x}}_{}   
  + 
    w^{\dagger}_\rmii{\vec{k}}\, e^{i k \tau - i \vec{k}\cdot\vec{x}}_{}   
 \, \Bigr]
 \;. \la{varphi_quantized_past}
\ee

We now go to de Sitter spacetime,\footnote{%
 It is straightforward to generalize the computation to first order 
 in slow-roll parameters~\cite{cr}, keeping track of a non-zero 
 value of $\dot{H}/H^2$, however then the mode 
 equations are solved by Bessel functions. As their properties are  
 intransparent and the conclusions remain unchanged, 
 we stick to pure de Sitter spacetime here.  
 } 
with the scale factor 
$
 a(t) = a( t^{ }_\rmi{ref} )\, e^{H ( t - t^{ }_\rmii{ref} ) }_{ }
$.
Denoting 
$
 a'_{ } \equiv {\rm d}a/{\rm d}\tau
$
and
$
 \dot{a} \equiv {\rm d}a/{\rm d}t
$,  
the scale factor satisfies
\be
 a' = 
 \dot a \,
 \frac{{\rm d}t}{{\rm d}\tau} = H a^2 
 \;. 
\ee
Choosing the range of $\tau$ as $(-\infty,0)$ and denoting $\H \equiv a'/a$, 
this leads to  
\be
 a = -\frac{1}{H \tau}
 \;, \quad
 \H = - \frac{1}{\tau} 
 \;. \la{a_eta}
\ee

The massless field equation 
in de Sitter spacetime reads
\be
 \scalar''_{ } + 2 \H \scalar'_{ }
 - \nabla^2 \scalar \; = \; 0
 \;. \la{varphi_eq_dS}
\ee
We stress that, contrary to a proper inflationary analysis
of scalar perturbations, 
there is no metric perturbation here 
to which $\scalar$ couples. 
Going over to co-moving momentum space, with 
\be
 \scalar = 
 \int \! \frac{{\rm d}^3\vec{k}}{(2\pi)^{3/2}_{ }} \, 
 \Bigl[\, 
 w^{ }_\rmii{\vec{k}}
 \, \scalar^{ }_k(\tau)
 e^{i \vec{k}\cdot\vec{x}}_{ }
 + 
 w^{\dagger}_\rmii{\vec{k}}
 \, \scalar^{*}_k(\tau)
 e^{-i \vec{k}\cdot\vec{x}}_{ }
 \,\Bigr]
 \;, \la{onsh}
\ee 
and inserting $\H$ from \eq\nr{a_eta}, 
the mode function $\scalar^{ }_k$ must fulfil 
\be
 \biggl(\,
    \partial_\tau^2 - \frac{2}{\tau}\, \partial^{ }_\tau + k^2
 \,\biggr) \, \scalar^{ }_k \; = \; 0
 \;. \la{varphi_eq_k}
\ee
This is readily solved, with 
the two independent solutions having the forms 
\be
 \scalar^{ }_k 
 \; \subset \; 
 \bigl\{\, 
 C\, (1 + i k \tau)\, e^{-i k \tau}
 \;, \quad
 C^*_{ } (1 - i k \tau)\, e^{i k \tau}
 \,\bigr\} 
 \;. \la{mode_funcs}
\ee
The constant $C$ is fixed so as to 
reproduce \eq\nr{varphi_quantized_past} at $\tau\to -\infty$, implying 
$
 C = i H / \sqrt{2 k^3}
$.
Thereby the full solution reads 
\be
 \scalar 
 = 2 H 
 \int \! \frac{{\rm d}^3\vec{k}}{ (4\pi k)^{3/2}_{ } } \, 
 \Bigl[ \,
    \bigl( i - k \tau \bigr)
    w^{ }_\rmii{\vec{k}}\, e^{-i k \tau + i \vec{k}\cdot\vec{x}}_{}   
  - \bigl( i + k \tau \bigr)
    w^{\dagger}_\rmii{\vec{k}}\, e^{i k \tau - i \vec{k}\cdot\vec{x}}_{}   
 \, \Bigr]
 \;. \la{varphi_quantized_now}
\ee

As a final step, we consider the equal-time correlation function.
Making use of \eq\nr{commutator}, \eq\nr{varphi_quantized_now} yields 
\be
 \langle 0 | \, 
 \scalar(\tau,\vec{x}^{ }_1) \, 
 \scalar(\tau,\vec{x}^{ }_2) \, 
 | 0 \rangle 
 \; = \; 
 \int \! \frac{{\rm d}^3\vec{k}}{(2\pi)^3} \, 
 e^{ i \vec{k}\cdot(\vec{x}^{ }_1 - \vec{x}^{ }_2 )}_{ }
 \, \frac{H^2(1 + k^2\tau^2)}{2 k^3}
 \;. \la{scalar_corr_dS}
\ee
The corresponding power spectrum, 
denoted by 
$
  \P^{ }_{\scalar}
$, 
is obtained by writing 
the integration measure as 
\be
 \frac{{\rm d}^3\vec{k}}{(2\pi)^3} 
 \; = \; 
 \frac{k^3 \, {\rm d}\ln k}{2\pi^2}
 \;, \la{measure}
\ee
and multiplying $k^3 / (2\pi^2)$ into the integrand of 
\eq\nr{scalar_corr_dS}. This results in 
\be
 \P^{ }_{\scalar}
 \; = \; 
 \biggl( \frac{H}{2\pi} \biggr)^2_{ }
 \bigl(\, 1 + k^2\tau^2 \,\bigr)
 \; \stackrel{\nr{a_eta}}{=} \; 
 \biggl( \frac{H}{2\pi} \biggr)^2_{ }
 \biggl(\, 1 + \frac{k^2}{a^2 H^2} \,\biggr)
 \;, \la{P_scalar}
\ee
which for $k \ll a H $
reduces to the text-book expression. 

%
\subsubsection{Stochastic derivation}
\la{sss:stochastic}

We proceed to repeat the computation in \se\ref{sss:canonical}
with another method~\cite{stochastic}, following
the presentation in ref.~\cite{rr}. 
Even if the 
intermediate steps look quite different, both technically and 
conceptually, \eq\nr{P_scalar} can indeed be reproduced. 
As mentioned, the reason for this exercise is that it helps
us to include thermal fluctuations 
in \ses\ref{ss:thermal} and \ref{ss:both}. 

The starting point is to divide $\scalar$ into short-distance
fluctuations ($\scalar^{ }_{<}$) and a slowly varying
long-distance part ($\scalar^{ }_{>}$), by writing 
\ba
 \scalar 
 & = & \scalar^{ }_{>} + \scalar^{ }_{<}
 \;, \la{varphi_splitup0} \\[2mm]
 \scalar^{ }_{<} 
 & \equiv & 
 \int \! \frac{{\rm d}^3\vec{k}}{\sqrt{(2\pi)^{3}_{ }}} \, 
 W(\tau,k) \, 
 \Bigl[ \,
    w^{ }_\rmii{\vec{k}}
    \, \scalar^{ }_k(\tau) 
    \, e^{ i \vec{k}\cdot\vec{x}}_{}   
  + 
    w^{\dagger}_\rmii{\vec{k}}
    \, \scalar^{*}_k(\tau) 
    \, e^{ - i \vec{k}\cdot\vec{x}}_{}   
 \, \Bigr]
 \;, \la{varphi_splitup}
\ea
where 
the mode functions $\scalar^{ }_k$ and $\scalar^*_k$ are from 
\eq\nr{mode_funcs}, and the window function $W$ can be chosen as
\be
 W(\tau,k) \; \equiv \; \theta(k - \epsilon a H)
 \;. \la{W}
\ee
The parameter $\epsilon$ drops out at the end. 
Inserting \eq\nr{varphi_splitup} into \eq\nr{varphi_eq_dS}, we get
\be
 \scalar''_{>} - \frac{2}{\tau}\, \scalar'_{>}
 - \nabla^2 \scalar^{ }_{>} \; = \; \xiQ^{ }
 \;, \quad
 \xiQ^{ } \; \equiv \; 
 - \biggl( 
       \partial_\tau^2 - \frac{2}{\tau}\, \partial^{ }_\tau - \nabla^2
   \biggr)
  \, \scalar^{ }_{<}
 \;. \la{varphi_eq_splitup}
\ee
The subscript in $\xiQ^{ }$ stands for quantum noise. 

If the differential operator in $\xiQ^{ }$ acts on the mode
functions $\scalar^{ }_k$ and $\scalar^*_k$ 
in $\scalar^{ }_{<}$, cf.\ \eq\nr{varphi_splitup}, 
then the result vanishes, just because
this is how the mode functions have been defined. Left over are terms acting
on the window function, 
\ba
 \xiQ^{ }(\tau,\vec{x}) & = & 
 - \int \! \frac{{\rm d}^3\vec{k}}{\sqrt{(2\pi)^{3}_{ }}} \, 
 \Bigl[ \,
    w^{ }_\rmii{\vec{k}}
    \, f^{ }_k(\tau) 
    \, e^{ i \vec{k}\cdot\vec{x}}_{}   
  + 
    w^{\dagger}_\rmii{\vec{k}}
    \, f^{*}_k(\tau) 
    \, e^{ - i \vec{k}\cdot\vec{x}}_{}   
 \, \Bigr]
 \;, \la{xi} \\[2mm]
 f^{ }_k(\tau) & \equiv & 
 \biggl( W''_{ } - \frac{2}{\tau}\, W'_{ } \biggr)
 \scalar^{ }_k 
 + 2 W'_{ } \scalar'_k
 \;. \la{f_k} 
\ea

Given that $W$ as defined by \eq\nr{W} is a step function, 
the terms in \eq\nr{f_k} are Dirac-$\delta$'s or
their derivatives. Concretely, recalling from \eq\nr{a_eta} that 
$a H = -1/\tau$, we can write $W = \theta(k + \epsilon/\tau)$. 
A consequence from here is that 
$\xiQ^{ }(\tau^{ }_1,\vec{x})$ 
and $\xiQ^{ }(\tau^{ }_2,\vec{x})$ 
are commuting (i.e.\ classical) variables
for $\tau^{ }_1 \neq \tau^{ }_2$: non-zero
contributions could originate from different momenta, 
$k^{ }_1 \neq k^{ }_2$, 
but then the operators 
$w^{ }_{\vec{k}^{ }_1}$ and $w^{\dagger}_{\vec{k}^{ }_2}$ commute, 
cf.\ \eq\nr{commutator}. However, the noise 
represented by~$ \xiQ^{ } $ is not obviously ``white'', 
as can be seen by determining its autocorrelator.
Evaluating this in the distant-past vacuum state, we get
\be
 \langle 0 | \, 
 \xiQ^{ }(\tau^{ }_1,\vec{x}^{ }_1) \, 
 \xiQ^{ }(\tau^{ }_2,\vec{x}^{ }_2) 
 \, | 0 \rangle 
 \; = \; 
 \int \! \frac{{\rm d}^3\vec{k}}{(2\pi)^3} \, 
 e^{i\vec{k}\cdot(\vec{x}^{ }_1 - \vec{x}^{ }_2)} 
 \, 
 f^{ }_k (\tau^{ }_1) \, f^*_k (\tau^{ }_2) 
 \;. \la{noise_qm}
\ee
This is not easily simplified, 
given the singular nature of the structures appearing, 
but in practice there is no need to take further steps.

It is convenient to go to spatial
momentum space,\footnote{%
 When we go to momentum space only in spatial directions 
 and consider quantities which do not satisfy a classical 
 equation of motion, the conventions are 
 $
  \xiQ^{ }(\tau^{ },\vec{k}) = 
  \int_\vec{x} e^{-i \vec{k}\cdot\vec{x}}
  \xiQ^{ }(\tau^{ },\vec{x})
 $ 
 and
 $
  \xiQ^{ }(\tau^{ },\vec{x}) = 
  \int_\vec{k} e^{i \vec{k}\cdot\vec{x}}
  \xiQ^{ }(\tau^{ },\vec{k})
 $, 
 with 
 $
  \int_\vec{k} \equiv 
  \int\! \frac{{\rm d}^3\vec{k}}{(2\pi)^3}
 $.
 This is to be contrasted with the representation 
 of on-shell fields, like in \eq\nr{onsh},
 where a definite time dependence appears in mode functions
 and the momentum integration measure is different.  
 } 
\be
 \langle 0 | \, 
 \xiQ^{ }(\tau^{ }_1,\vec{k}) \, 
 \xiQ^{ }(\tau^{ }_2,\vec{q}) 
 \, | 0 \rangle 
 \; = \; 
 \deltabar(\vec{k+q})
 \, 
 f^{ }_k (\tau^{ }_1) \, f^*_k (\tau^{ }_2) 
 \;. \la{noise_qm_k}
\ee
where
$
 \int_\vec{k} \;\deltabar(\vec{k}) \equiv 1
$.
Equation~\nr{varphi_eq_splitup} then becomes
\be
 \biggl(\, 
  \partial_\tau^2 - \frac{2}{\tau}\, \partial^{ }_\tau + k^2
 \,\biggr) \, 
 \scalar^{ }_{>}(\tau,\vec{k})
 = \xiQ^{ }(\tau,\vec{k})
 \;. \la{varphi_eq_splitup_k}
\ee

Let us solve \eq\nr{varphi_eq_splitup_k} with 
a retarded Green's function, 
denoted by $G^{ }_\R(\tau,\taui^{ },k)$. 
It satisfies 
\be
 \biggl(\, 
  \partial_\tau^2 - \frac{2}{\tau}\, \partial^{ }_\tau + k^2
 \,\biggr) \, 
 G^{ }_\R(\tau,\taui^{ },k) = 
 \delta(\tau - \taui^{ }) 
 \;, \la{G_R}
\ee
with the boundary condition 
$
 G^{ }_\R(\tau,\taui^{ },k) 
 \stackrel{ \tau < \taui^{ } }{\equiv} 0  
$. 
For $\tau > \taui^{ }$, the solution is a linear combination of 
the mode functions in \eq\nr{mode_funcs}. The coefficients 
are obtained by integrating \eq\nr{G_R} over the source, implying
\be
 \lim^{ }_{\tau\to\taui^{+}}
 G^{ }_\R(\tau,\taui^{ },k) = 0 
 \;, \quad
 \lim^{ }_{\tau\to\taui^{+}}
 \partial^{ }_\tau 
 G^{ }_\R(\tau,\taui^{ },k) = 1
 \;. \la{G_R_bc}
\ee
The explicit expression reads 
\be
 G^{ }_\R(\tau,\taui^{ },k)
 \; = \; 
 \frac{
 \theta(\tau - \taui^{ })
   }{k^3 \taui^2}
 \, 
   \im\bigl[\, 
   e^{ ik(\tau - \taui^{ })}_{ } (1 - i k \tau)(1 + i k \taui^{ })
   \,\bigr]
 \;, \la{G_R_soln}
\ee
and the solution becomes
\be
 \scalar^{ }_{>}(\tau,\vec{k}) 
 = 
 \int_{-\infty}^{\tau} \, {\rm d}\taui^{ } \, 
 G^{ }_\R(\tau,\taui^{ },k) \, \xiQ^{ }(\taui^{ },\vec{k})
 \;. 
 \la{varphi_eq_soln_k}
\ee

We now consider the equal-time correlator.
In momentum space, it reads 
\ba
 \langle\, 
 \scalar^{ }_{>}(\tau,\vec{k}) \, 
 \scalar^{ }_{>}(\tau,\vec{q}) 
 \,\rangle
 & = & 
 \int_{-\infty}^{\tau} \!\!\! {\rm d}\tau^{ }_1 \, 
 G^{ }_\R(\tau,\tau^{ }_1,k)
 \int_{-\infty}^{\tau} \!\!\! {\rm d}\tau^{ }_2 \, 
 G^{ }_\R(\tau,\tau^{ }_2,q)
 \, 
 \langle 0 | \, 
 \xiQ^{ }(\tau^{ }_1,\vec{k}) \, 
 \xiQ^{ }(\tau^{ }_2,\vec{q}) 
 \, | 0 \rangle 
 \nn 
 & \!\!\! \stackrel{\rmii{\nr{noise_qm_k}}}{=} \!\!\! & \;\;
 \deltabar(\vec{k+q})
 \, 
 \biggl| 
  \int_{-\infty}^{\tau} \, {\rm d}\taui^{ } \, 
  G^{ }_\R(\tau,\taui^{ },k) \, f^{ }_k(\taui^{ })
 \biggr|^2_{ }
 \;. \la{stoch_2pt}
\ea
Inserting \eq\nr{f_k} and 
carrying out a partial integration, we find
\ba
 && \hspace*{-1.3cm} 
  \int_{-\infty}^{\tau} \, {\rm d}\taui^{ } \, 
  G^{ }_\R(\tau,\taui^{ },k) \, f^{ }_k(\taui^{ })
 \la{G_R_full} \\ 
 & = & 
  \int_{-\infty}^{\tau} \, {\rm d}\taui^{ } \, W'(\taui^{ }) \, 
 \biggl\{ 
   - \partial^{ }_{\taui^{ }} G^{ }_\R(\tau,\taui^{ },k)
    \scalar^{ }_k(\taui^{ }) 
   + G^{ }_\R(\tau,\taui^{ },k)
   \biggl[
    \scalar'_k(\taui^{ }) -  
    \frac{2}{\taui^{ }}\, \scalar^{ }_k(\taui^{ }) 
   \biggr]
 \biggr\}
 \;. \nonumber  
\ea
Making use of $a H  = -1/\tau$, 
we can write 
$
 W'_{ }(\taui^{ }) = 
 \delta(k + \frac{\epsilon}{\taui^{ }}) \frac{k}{\taui^{ }}
 = 
 \delta(\taui - \tau^{ }_{*k})
$, 
where 
$\tau^{ }_{*k}$ is the time at which the momentum mode $k$ 
crosses the horizon. This then leads to  
\be
 \int_{-\infty}^{\tau} \! {\rm d}\taui^{ } \, 
  W'_{ }(\taui^{ }) 
  \, \mathcal{F}(\taui^{ })
  \; = \; 
  \mathcal{F}(\tau^{ }_{*k})
 \;. \la{W_int}
\ee

It remains to insert into \eqs\nr{G_R_full} and \nr{W_int}
the mode function $\scalar^{ }_k$ from
\eq\nr{mode_funcs}, 
its time derivative~$\scalar'_k$, 
as well as the Green's function from 
\eq\nr{G_R_soln}. There is a remarkable cancellation
between many terms, so that any dependence on $\tau^{ }_{*k}$
drops out. We find 
\be
  \int_{-\infty}^{\tau} \, {\rm d}\taui^{ } \, 
  G^{ }_\R(\tau,\taui^{ },k) \, f^{ }_k(\taui^{ })
 \; = \; 
 \frac{i H}{\sqrt{2k^3}} 
 \bigl( 1 + i k \tau\bigr) e^{-i k \tau}_{ }
 \; \stackrel{!}{=} \; 
 \scalar^{ }_k(\tau)
 \;. \la{memory_int}
\ee
Taking the absolute value squared, as needed in \eq\nr{stoch_2pt},
and multiplying by $k^3/(2\pi^2)$, like between 
\eqs\nr{measure} and \nr{P_scalar}, 
the power spectrum in \eq\nr{P_scalar} is recovered.   

%
\subsubsection{Including helicity}
\la{sss:helicity}

The helicity components of 
tensor perturbations of the metric, or gravitational waves, satisfy 
the same equation in de Sitter spacetime as 
the scalar fluctuations considered in 
\ses\ref{sss:canonical} and \ref{sss:stochastic}. 
Thus, once we fix normalization, 
via a comparison with \eq\nr{normalization}, 
and take care of the sum over helicities or polarizations, 
the results can be extracted from above. 
We denote tensor perturbations of the metric 
by $\delta {g^{\rmi{t}\hspace*{0.2mm}i}_{ }}^{ }_j \equiv h^\rmi{t}_{ij}$, 
with the convention that spatial indices are raised and lowered
like in flat spacetime, 
i.e.\ 
$
 T^{ }_{ij} \equiv T^{ij}_{ }\equiv {T^i_{ }}^{ }_j
$.

In a local Minkowskian frame, 
the energy associated with tensor perturbations reads
\be
 \bigl[\, E^{ }_\rmiii{GW} \,\bigr]^{ }_\M = 
 \frac{1}{32\pi G} 
 \int \! {\rm d}^3\vec{y} \, 
 ( \dot{h}^{\rmi{t}}_{ij} )^2_{ }
 \;. \la{E_GW_1}
\ee
If we express a wave
as a linear combination of forward and backward propagating modes, 
and integrate over a time larger than the oscillation period, 
so that fast oscillations $e^{\pm 2 i p t}_{ }$ average out, 
then \eq\nr{E_GW_1} can be substituted through
\be
 \bigl\langle\, 
   E^{ }_\rmiii{GW}
 \,\bigr\rangle^{ }_\M = 
 \frac{1}{32\pi G} 
 \int \! {\rm d}^3\vec{y} \, 
 \frac{1}{2} 
  \bigl[ \, 
   (\dot{h}^{\rmi{t}}_{ij})^2_{ } 
   + 
   (\nabla h^\rmi{t}_{ij})^2_{ }
  \, \bigr]
 \;. \la{E_GW_2}
\ee
Comparing with \eq\nr{normalization}, 
canonically normalized tensor modes, 
denoted in the following 
by~$ 
 \hat{h}^\rmi{t}_{ij}
$,  
are obtained as 
$
 \hat{h}^\rmi{t}_{ij} \equiv 
 h^\rmi{t}_{ij} / \sqrt{32\pi G}
$.

There are two propagating helicity states in the tensor channel. 
Denoting the normalized and complete set of polarization tensors by
$
  \epsilon^{\lambda}_{ij}
$, 
we may write in momentum space
\be
  \hat{h}^\rmi{t}_{ij}
  \; = \;  
  \sum_\lambda 
  \epsilon^{\lambda}_{ij}\,
  \hat{h}^{\lambda}_{ }
  \;, \quad
  \sum_{i,j}
  \epsilon^{\lambda}_{ij}\,
  \epsilon^{\lambda'*}_{ij}
 \; = \; 
 \delta^{\lambda\lambda'}_{ }
 \;. \la{helicity_1}
\ee
For the same momentum, 
each state obeys the same equation and carries the same energy density. 
Therefore the power spectrum 
$
 \P^{ }_\rmii{T}
$, 
associated with 
$
 \sum_{\lambda} h^{\lambda}_{ } 
$, 
becomes
\be
 \P^{ }_\rmii{T}(k)
 \; = \; 
 \underbrace{\; 32 \pi  G \;}_{h^2 / \hat{h}^2} 
 \times 
    \underbrace{\! 2 \!}_{\sum_\lambda}
 \times 
 \underbrace{
 \biggl( \frac{H}{2\pi} \biggr)^2_{ }
 \biggl(\, 1 + \frac{k^2}{a^2 H^2} \,\biggr)}_{\rmii{\nr{P_scalar}}}
 \; = \; 
 \frac{16}{\pi}
 \biggl( \frac{H}{\mpl^{ }} \biggr)^2_{ }
 \biggl(\, 1 + \frac{k^2}{a^2 H^2} \,\biggr)
 \;. \la{P_T}
\ee

As for the practical use of \eq\nr{P_T}, we recall that during
inflation, $H$ is approximately constant whereas $a$ grows rapidly
($k$ is constant by definition). Conventionally,  
$
  \P^{ }_\rmii{T}(k)
$
is evaluated at horizon crossing, with $k = a H$. Taken literally, 
the second term in \eq\nr{P_T} would then represent a 100\% correction, 
but this is an illusion, as a short moment later it would be minuscule,
whereas the first term would remain intact. 
Therefore, we can adopt the standard convention $k = a H$ 
for horizon crossing, 
with the implicit understanding that 
the second term in \eq\nr{P_T} needs to be omitted, 
recovering the standard literature result. 

%
\subsection{Thermal fluctuations}
\la{ss:thermal}

Apart from vacuum fluctuations, tensor modes can also be produced
by thermal ones, via the Einstein equation 
\be
 \bigl(\,
   \partial^2_\tau 
    + 2 \H \,\partial^{ }_\tau
    - \nabla^2 
 \,\bigr) h^\rmi{t}_{ij}
 = 
 16 \pi G a^2 \, T^\rmi{t}_{ij}
 \;, \la{tensor_thermal}
\ee
where $ T^\rmi{t}_{ij} $ is the tensor part of 
a matter energy-momentum tensor. Even if the average value
of $ T^\rmi{t}_{ij} $ vanishes, non-zero values are 
produced by fluctuations. 
At large wavelengths (small momenta), the
latter are known as hydrodynamic fluctuations. 
Though this is in principle
a text-book topic, the treatment of hydrodynamic fluctuations  
often does not belong to our standard tool kit.
It turns out that for the tensor modes, the story is 
simpler than for scalar and vector modes, and we recall here
the basic ingredients. 

In the hydrodynamic domain, the energy-momentum tensor is 
expressed in terms of a gradient expansion operating on macroscopic 
variables, namely the local temperature, $T$, 
and the local fluid velocity, $v^{ }_i$.
The zeroth order contains no gradients, and is known 
as ideal hydrodynamics, with the dependence on $T$ parametrized
through a local energy density and pressure ($e$ and $p$, respectively). 
At first order in gradients, the dependence on $T$ is parametrized
by two new quantities, the shear and bulk viscosities
($\eta$ and $\zeta$, respectively). Crucially, 
at the same order, 
hydrodynamic fluctuations need to be included~\cite{landau9}. 
The underlying reason is the 
fluctuation-dissipation theorem: viscosities are dissipative
coefficients, which transfer energy from fluid motion to thermal 
noise. This must be compensated for by fluctuations, 
returning energy from thermal noise to the fluid degrees of freedom. 

Concretely, at first order in gradients, the energy-momentum tensor
can be expressed as 
\be
 T^{ \mu\nu}_\rmi{hydro} = \bar{T}^{\mu\nu}_\rmi{hydro} + 
                   \delta T^{\mu\nu}_\rmi{hydro} + 
                   S^{\mu\nu}_{ } + \rmO(\delta^2)
 \;. \la{Tmunu_expansion}
\ee
Here $ \bar{T}^{\mu\nu}_\rmi{hydro} $ contains 
the average values of the hydrodynamic variables 
($T\equiv \bar T, \bar{v}^{ }_i \equiv 0$), 
$\delta T^{\mu\nu}_\rmi{hydro}$ their expansions to first order
($\delta T, v^{ }_i \equiv \delta v^{ }_i $), and 
$S^{\mu\nu}_{ }$ the fluctuations. 
At first order, recalling our convention that spatial indices 
are raised and lowered with $\delta^{ij}_{ }$ and $\delta^{ }_{ij}$, 
the fluctuation autocorrelator has the form~\cite{landau9,kapusta}
\be
 \langle\, S^{ij}_{ }(\X)\, S^{mn}_{ }(\Y) \,\rangle
 = 
 2 T \, 
 \biggl[
 \eta \, \bigl( 
                 \delta^{ }_{im} \delta^{ }_{jn}
               + \delta^{ }_{in} \delta^{ }_{jm}
         \bigr)
 +       \biggl( 
                 \zeta - \frac{2\eta}{3}
         \biggr) \, 
                 \delta^{ }_{ij} \delta^{ }_{mn} 
 \biggr] 
 \, \frac{\delta_{ }(\mathcal{X-Y})}
 {\sqrt{- \det g^{ }_{\mu\nu}}}
 \;. \la{S_S_0}
\ee
Only the spatial part displays non-vanishing correlators at this order. 

Now, when we consider the production of tensor perturbations, we go to 
momentum space in the spatial directions, and sum over the graviton 
polarization states. With the notation from \eq\nr{helicity_1}, the
latter produces
\ba
 \sum_{\lambda}
  \epsilon^{\lambda}_{ij}\,
  \epsilon^{\lambda*}_{mn}
 & = & 
  \mathbbm{L}^{ }_{ij;mn}
 \;, \la{helicity_2} \\
 \mathbbm{L}^{ }_{ij;mn} 
 & \equiv &
 \frac{ 
      \PT_{im}\PT_{jn} 
   +  \PT_{in}\PT_{jm}
   -  \PT_{ij}\PT_{mn}
      }{2} 
 \;, \quad
 \PT_{ij}
 \; \equiv \; 
 \biggl(
   \delta^{ }_{ij} - \frac{k^{ }_i k^{ }_j}{k^2} 
 \biggr)
 \;. \la{def_L}
\ea
The 2-point function of tensor fluctuations of the energy-momentum tensor
thus contains 
\ba
 && \hspace*{-1.5cm}
 \mathbbm{L}^{ }_{ij;mn}
 \Bigl\langle\, 
 T^{ij}_\rmi{hydro}(\tau^{ }_1,\vec{k}) 
 T^{mn}_\rmi{hydro}(\tau^{ }_2,\vec{q}) 
 \,\Bigr\rangle
 \nn 
 & = & 
 \mathbbm{L}^{ }_{ij;mn} \, 
 \Bigl\langle \, 
 [ \delta T^{ij}_\rmi{hydro} + S^{ij}_{ }]
   (\tau^{ }_1,\vec{k}) \,
 [ \delta T^{mn}_\rmi{hydro} + S^{mn}_{ }]
   (\tau^{ }_2,\vec{q})
 \, \Bigr\rangle 
 \nn 
 & = & 
 \mathbbm{L}^{ }_{ij;mn} \, 
 \Bigl\langle \, 
   S^{ij}_{ }
   (\tau^{ }_1,\vec{k}) \,
   S^{mn}_{ }
   (\tau^{ }_2,\vec{q})
 \, \Bigr\rangle 
 \nn[2mm] 
 & = & 
 \frac{ 8 T \eta \, 
 \; \deltabar(\vec{k+q}) 
 \, \delta(\tau^{ }_1 - \tau^{ }_2) 
 }{a^4}  
 \;. \la{Tij_tensor_0}
\ea
In the penultimate step we made use of the fact that 
$ \delta T^{ij}_\rmi{hydro} $ 
has no projection onto tensor modes, so 
that the correlator arises directly from the fluctuations. 
Let us remark that the projection could be carried
out without a helicity sum as well, yielding then
$
 8 T \eta \to 4 T \eta\, \delta^{\lambda\lambda'}_{}
$.

In order to make use of \eq\nr{Tij_tensor_0}, we need to 
know the value of the shear viscosity. 
In a system with a thermalized inflaton coupled to a heat bath 
through a damping coefficient~$\Upsilon$, the value of 
the shear viscosity emanating from $\Upsilon$ can be obtained 
from \eq(3.19) of ref.~\cite{gravity}, 
by converting the imaginary part of the retarded correlator to 
a statistical (time-symmetric) correlator and taking the IR limit, 
\be
 T \eta 
 \; \supset \;
 T \eta^{ }_\chi
 \; \equiv \; 
 \lim_{\omega,k\to 0} 
 \frac{T  \im G^\R_{xy;xy} (\omega,k) }{\omega}
 \; = \; 
 \int^{ }_\vec{p} 
 \frac{ p_x^2 p_y^2 
 \,\nB^{ }(\epsilon^{ }_p)
 \bigl[
  1 +  \nB^{ }(\epsilon^{ }_p) 
 \bigr]
      }{ {\Upsilon} \, {\epsilon}_p^2} 
 \;, \quad
 \epsilon_p^2 \equiv p^2 + m^2
 \;. \la{eta_varphi}
\ee
This shows how a weak coupling leads to a large
shear viscosity, a phenomenon that we can physically associate
with a long mean free path. However, the shear viscosity gets
contributions also from reactions not involving the inflaton; 
in a pure gauge plasma, it reads~\cite{amy2}
\be
 T \eta 
 \; \supset \;
 T \eta^{ }_\rmi{gauge}
 \; 
 \underset{ \rmii{$ \Nf = 0 $ } }{  
  \overset{ \rmii{$ \Nc = 3 $ } }{ \simeq } } 
 \; 
 \frac{27.126\,T^4}{g^4 \ln \bigl( \frac{2.765\,T}{m^{ }_\rmiii{D}} \bigr)}
 \;, \quad
 m^{2}_\rmii{D} = \frac{g^2 T^2}{3}
 \;. \la{eta_gauge}
\ee
For $T \gg m$, \eq\nr{eta_varphi} dominates, as 
$\Upsilon \ll \epsilon^{ }_p$ can be very small compared with $T$; 
for $T \ll m$, 
\eq\nr{eta_gauge} turns into the leading contribution,  
given that \eq\nr{eta_varphi} is suppressed by $e^{-m/T}$. 

%
\subsection{Combining vacuum and thermal fluctuations}
\la{ss:both}

In the warm inflation literature, 
it is often stated that there are no thermal 
corrections to tensor perturbations, unless gravitons
get thermalized, which is unlikely (see below). 
That said, thermal corrections were addressed in ref.~\cite{sorbo},
and argued to be possibly important; 
at the same time, 
the vacuum part was omitted from the formalism, 
even if at the end the thermal contribution was 
compared with it. The approach below 
largely agrees with that of ref.~\cite{sorbo}, 
but incorporates the vacuum part all along and does not
assume de Sitter spacetime. 

Let us first briefly contrast with scalar perturbations. 
These are damped by 
the coefficient~$\Upsilon$ appearing in \eq\nr{eta_varphi}, 
and excited by thermal noise~\cite{hydro}. In principle, 
tensor perturbations also experience damping, 
with an associated coefficient~$\Upsilon^\rmi{t}_{ }$.
However,  
for physical momenta $ k/a \lsim \alpha^2 T $, 
$\Upsilon^\rmi{t}_{ }$
is proportional to the shear viscosity $\eta$, 
divided by $\mpl^2$, 
and in total of order 
$
 \Upsilon^\rmi{t}_{ } \sim 
 T^3/(\alpha^2 \mpl^2) 
$~\cite{gravity_qualitative}. 
This is normally smaller than the Hubble rate,  
$
 \Upsilon^\rmi{t}_{ } \ll H \sim T^2/\mpl^{ } 
$. 
Therefore gravitons are not damped or thermalized: 
$
 \Upsilon^{t}_{ }
$
can be omitted, 
and we associate no Bose distribution to gravitons.

Going to momentum space, 
multiplying \eq\nr{tensor_thermal} with the polarization vector
$\epsilon^\lambda_{ij}$, 
and summing over the spatial indices, 
we find
\be
 \bigl(\,
   \partial^2_\tau 
     + 2 \H \,\partial^{ }_\tau
     + k^2 
 \,\bigr)
 h^\lambda_{ }(\tau,\vec{k})
 \; = \; 
 16 \pi G a^2 \, \sum_{i,j} \epsilon^\lambda_{ij}
 T^\rmi{t}_{ij}(\tau,\vec{k})
 \; \equiv \; 
 \xiT^{\lambda}(\tau,\vec{k})
 \;. \la{tensor_T}
\ee
Dividing $h^\lambda_{ }$ into 
short-distance fluctuations and a slowly varying part, 
like in \eq\nr{varphi_splitup0}, 
the evolution equation for the latter 
can then be expressed as 
\be
 \bigl(\, 
  \partial_\tau^2 + 
   2 \H \, \partial^{ }_\tau
  + k^2 
 \,\bigr) 
 h^{\lambda}_{>}(\tau,\vec{k})
 \; = \; 
 \bigl( \, 
    \xiQ^{\lambda}  + \xiT^{\lambda} 
 \, \bigr) (\tau,\vec{k})
 \;. \la{master_T}
\ee
Recalling the connection of $h^\lambda_{ }$ to canonically 
normalized fields, like below \eq\nr{E_GW_2}, 
the autocorrelator 
of the vacuum fluctuations 
in \eq\nr{noise_qm_k} takes the form
\ba
 \bigl\langle\, 
   \xiQ^{\lambda}(\tau^{ }_1,\vec{k}) \, 
   \xiQ^{\lambda'}(\tau^{ }_2,\vec{q})
 \,\bigr\rangle
 & = & 
  \; \deltabar(\vec{k+q}) \, \delta^{\lambda\lambda'}_{ } 
  f^{ }_k (\tau^{ }_1) \, f^*_k (\tau^{ }_2) 
  \, 32\pi G
 \;. \la{tensor_noise_Q} 
\ea
For the thermal noise autocorrelator, the definition in 
\eq\nr{tensor_T} together with the hydrodynamic noise in 
\eq\nr{Tij_tensor_0} imply that 
\ba
 \bigl\langle\, 
   \xiT^{\lambda}(\tau^{ }_1,\vec{k}) \, 
   \xiT^{\lambda'}(\tau^{ }_2,\vec{q})
 \,\bigr\rangle
 & = & 
 \; \deltabar(\vec{k+q}) 
 \, \delta^{\lambda\lambda'}_{ }
 \, \delta(\tau^{ }_1 - \tau^{ }_2) 
 \, 4 T (\tau^{ }_1 )\, \eta(\tau^{ }_1) 
 \, (16\pi G)^2_{ }
 \;. \la{tensor_noise_T}
\ea
The mixed noise correlator vanishes, since the impact of gravitational waves
on the thermal plasma, or vice versa, is suppressed by higher powers
of~$G$.

Solving \eq\nr{master_T} with Green's functions,
cf.\ \eq\nr{G_R_soln}, and recalling that
there are two polarization states, 
the power spectrum at a time $\taue^{ }$ finally becomes
\be
  \P^{ }_\rmii{T}(k) = 
  \frac{64\pi G  \, k^3 }{2\pi^2}
  \, \biggl\{ 
  \biggl| 
   \int_{-\infty}^{\taue^{ }} \!\!\! {\rm d}\taui^{ } \, 
   G^{ }_\R(\taue^{ },\taui^{ },k) 
  \, f^{ }_k(\taui^{ })
  \biggr|^2_{ }
  \! + 32 \pi G \! 
    \int_{-\infty}^{\taue^{ }} \!\!\! {\rm d}\taui^{ } \, 
    G^{2}_\R(\taue^{ },\taui^{ },k) \, 
    T(\taui^{ })\, \eta(\taui^{ })
  \biggr\} 
  \;. \la{master_P_T}  
\ee
Here we have introduced the notation $\taue^{ }$ 
for some arbitrarily chosen moment shortly after inflation,
such that $k / (a^{ }_e H^{ }_e) \ll 1$ for the momenta that affect
phenomenological predictions today (cf.\ \se\ref{ss:eGW}).
In the slow-roll regime 
the first part of \eq\nr{master_P_T} can be simplified, 
cf.\ \eq\nr{memory_int}, but the second part not, because
$T$ and $\eta$ can be complicated functions of time, and 
because their dominant contribution can originate at a late
time when de Sitter is no longer a good approximation. 

We note from \eq\nr{master_P_T} that the thermal contribution
to the tensor power spectrum is suppressed by $1/\mpl^2$
compared with the vacuum contribution.
However, while the vacuum contribution is localized in time, 
originating when the modes cross the horizon, 
this is not necessarily the case 
for the thermal contribution, which can continue for a while,
thereby compensating for the apparent suppression. 

%
\subsection{From primordial spectrum 
to the current gravitational energy density}
\la{ss:eGW}

The current-day observable corresponding to 
$\P^{ }_\rmii{T}$ is the power spectrum of the 
fractional energy density carried by 
gravitational waves, denoted by $\Omega^{ }_\rmiii{GW}$ and 
defined in \eq\nr{Omega_GW}. 
We recall here how the transfer function from 
$\P^{ }_\rmii{T}$  to $\Omega^{ }_\rmiii{GW}$
can be obtained, 
following the presentation 
in refs.~\cite{eos0,eos2}
(early work can be found in ref.~\cite{djs}). 

The starting point is the energy in \eq\nr{E_GW_1}, averaged
like in \eq\nr{E_GW_2}. We simultaneously take a vacuum and thermal
average, like in \eqs\nr{tensor_noise_Q} and \nr{tensor_noise_T}. 
Furthermore we make use of translational invariance, dividing
by volume and thereby defining
the energy density. Going over to conformal time, 
the energy density today ($\tau \to \tau^{ }_0$, $a\to a^{ }_0$) reads
\be
 e^{ }_\rmiii{GW}(\tau^{ }_0) 
 = 
 \frac{1}{32\pi G a_0^2}
 \sum_{i,j} 
 \bigl\langle\,
   {h^{\rmi{t}}_{ij}\hspace*{-0.8mm}}' (\tau^{ }_0,\vec{y})\,
   {h^{\rmi{t}}_{ij}\hspace*{-0.8mm}}' (\tau^{ }_0,\vec{y})
 \,\bigr\rangle
 \;, \la{e_GW}
\ee
where the left-hand side does not depend on $\vec{y}$ because
of translational invariance. 
The energy density in \eq\nr{e_GW} can be 
compared with the total energy density today
(cf.\ \eq\nr{eom_friedmann}), 
\be 
 e^{ }_\rmi{crit} 
 \; \equiv \; 
 e^{ }_r(\tau^{ }_0) 
 \; = \; 
 \frac{3 H_0^2}{8 \pi G}
 \;, \la{e_crit} 
\ee
where the current Hubble rate is conventially expressed as  
$H^{ }_0 = 100\,h$~km~s$^{-1}_{ }$~Mpc$^{-1}_{ }$.

We now adopt co-moving Fourier space in spatial directions, 
and represent the tensor structure in the helicity basis, 
like in \eq\nr{helicity_1}. The tensor perturbation
is written as a functional\hspace*{0.2mm}\footnote{%
 Given that $h^{\lambda}_{ }$ obeys a second
 order differential equation, both its initial value and 
 time derivative 
 are needed for specifying the solution. The notation
 assumes that $\taue^{ }$ is chosen
 at a moment where the conformal time derivative vanishes, 
 because the modes are well outside the horizon. For simplicity
 we also envision that the system has reheated by this time, i.e.\
 that the energy density is dominated by radiation.  
 } 
of its initial value at the end of inflation, 
\be
 h^{\lambda}_{ }(\tau^{ }_0,k) \; \equiv \;  
 X(\tau^{ }_0,\taue^{ },k) \,
 h^{\lambda}_{ }(\taue^{ },k)
 \;, \quad
 X(\taue^{ },\taue^{ },k) \; \equiv \; 1
 \;, \quad
 \partial^{ }_{\tau^{ }_0} X(\tau^{ }_0,\taue^{ },k) 
 \; \stackrel{\tau^{ }_0 = \taue^{ }}{\equiv} \; 0
 \;. \la{def_X}
\ee
This leads to 
\be
 \Omega^{ }_\rmiii{GW}(k) 
 \; \equiv \; 
 \frac{1}{e^{ }_\rmi{crit}}
 \frac{{\rm d} e^{ }_\rmiii{GW}(\tau^{ }_0)}{{\rm d}\ln k}
 \; = \; 
 \frac{[\partial^{ }_{\tau^{ }_0}
  X(\tau^{ }_0,\taue^{ },k)]^2 \P^{ }_\rmii{T}(k) }
 {32\pi G a_0^2\, e^{ }_\rmi{crit}}
 \; = \; 
 \underbrace{ 
 \frac{[\partial^{ }_{\tau^{ }_0}
  X(\tau^{ }_0,\taue^{ },k)]^2  }
 {12 a_0^2 H_0^2}
 }_{ \; \equiv \; \mathcal{T}^{ }_\rmiii{T}(k)} 
 \, \P^{ }_\rmii{T}(k)
  \la{Omega_GW} 
 \;, 
\ee
where 
$
 \mathcal{T}^{ }_\rmii{T}(k)
$
is called the transfer function in the tensor channel.
Normally, 
$
 \Omega^{ }_\rmiii{GW} 
$
is expressed in terms of the current-day frequency, $f^{ }_0$, 
as
$
 \Omega^{ }_\rmiii{GW}(f^{ }_0) 
$.

Given that $ \mathcal{T}^{ }_\rmiii{T} $ 
accounts for physics after the reheating period, 
thermodynamic functions are dominated by their radiation parts. 
The average energy density, pressure, entropy density, and heat
capacity satisfy standard relations, 
$
 {s}^{ }_r = \partial^{ }_\T {p}^{ }_r
$, 
$
 {e}^{ }_r = T {s}^{ }_r - {p}^{ }_r
$, 
$
 {c}^{ }_r \equiv \partial^{ }_\T {e}^{ }_r = T \partial^{ }_\T {s}^{ }_r
$.
We also make use of the speed of sound squared, 
$
 c_s^2 \equiv \partial^{ }_\T {p}^{ }_r / 
              \partial^{ }_\T {e}^{ }_r = {s}^{ }_r / {c}^{ }_r
$.
Unlike for inflation, where an SU($\Nc^{ }$) gauge
plasma was assumed to constitute the heat bath, we include
the full Standard Model in these
post-inflationary thermodynamic functions.

For tracking the evolution of the temperature, 
it is helpful to employ the variable 
\be
 z \; \equiv \; \ln\biggl( \frac{T^{ }_{e}}{T} \biggr)
 \;,
 \la{def_z}
\ee
where $T^{ }_e$ denotes the temperature at the end of inflation.
In practice we have chosen 
$T^{ }_{e} \equiv 10^{-9}_{ }\mpl^{ } = 1.22091\times 10^{10}_{ }$~GeV, 
but this is just a convention and has no physical effect.

Now, the scale factor, $a$, can be expressed in terms of $z$. 
In the absence of $\bar\varphi$, 
the background equation, \eq\nr{eom_plasma}, 
can be written as
$
  e_r' = - 3 \H ({e}^{ }_r + {p}^{ }_r)
$, 
which after the insertion of the thermodynamic identities turns into 
\be
 \bigl(\, {s}^{ }_r a^3 \,\bigr)' = 0 
 \;. \la{entropy}
\ee
{}From here we get
\be
 a \; = \; a^{ }_e \biggl( \frac{s^{ }_e}{s^{ }_r} \biggr)^{1/3}_{ }
 \;, \quad
 \H = 
 \sqrt{\frac{8\pi e^{ }_r}{3}} \frac{a^{ }_e}{\mpl^{ }}
 \biggl( \frac{s^{ }_e}{s^{ }_r} \biggr)^{1/3}_{ }
 \;. \la{scale_factor}
\ee
In addition \eq\nr{entropy} gives an evolution equation for $z$ itself, 
\be
 \partial^{ }_u z = \frac{ 3 c_s^2\, \H}{k} 
 \;, \quad
 u \; \equiv \; k \tau
 \;, \la{dotT}
\ee
where we made use of thermodynamic identities
and introduced the helpful variable $u$~\cite{sw}.
Using \eq\nr{def_X} in \eq\nr{master_T}
and neglecting the right-hand side,\footnote{%
 The physics of ``neutrino free streaming''
 in a gravitational wave background after their decoupling
 (cf.\ ref.~\cite{sw} and references therein) 
 induces an additional term in the equation, 
 but it only affects frequencies
 $f^{ }_0 \lsim 10^{-9}_{ }$~Hz~\cite{eos0}.
 These are far below the LISA window and thus not relevant for us here. 
 }
$X$ in turn satisfies
\be
 \partial_u^2 {X} + \frac{ 2 \H \partial^{ }_u {X} }{k}  + X = 0
 \;, \quad
 u > u^{ }_e
 \;. \la{X_eq_1}
\ee

In order to solve \eqs\nr{dotT} and \nr{X_eq_1}, we need to know the 
evolution of the ratio $k / \H$. Co-moving momenta 
are conveniently parametrized by the current 
frequency, $f^{ }_0/$Hz. Let us express $k$ in terms of the physical 
momentum evaluated at present time, $p^{ }_0$,
as $ k = a^{ }_0\, p^{ }_0$. 
In natural units ($\hbar = c = k^{ }_\rmii{B} = 1$),
$p^{ }_0 = \omega^{ }_0 = 2 \pi f^{ }_0$. 
Making use of \eq\nr{entropy}, 
we may then write
\be
 \frac{k}{\H}
 \; = \; \frac{k}{aH}
 \; = \; \frac{a^{ }_0 p^{ }_0}{a H}
 \; = \; \frac{p^{ }_0}{T^{ }_0}
 \biggl( \frac{a_0^3 T_0^3}{a^3_{ }T^3_{ }} \biggr)^{\fr13}_{ }
 \frac{T}{H}
 \; = \; 
 \underbrace{ 
 \frac{2\pi}{ \mbox{s}\, T^{ }_0}
 \sqrt{\frac{3}{8\pi}} 
 }_{ 
 \approx\; 
 6.0837 \times 10^{-12}_{ } 
 }
 \biggl(
   \frac{{s}^{ }_r/T^3}{{s}^{ }_0 / T_0^3} 
 \biggr)^{\fr13}_{ }
 \frac{\mpl^{ }/T}{\sqrt{{e}^{ }_r/T^4}}\,
 \frac{f^{ }_0}{\mbox{Hz}}
 \;. \la{k_aH_1}
\ee
Tabulated results for $c_s^2$ (for \eq\nr{dotT}) as well as
${s}^{ }_r/T^3_{ }$ and ${e}^{ }_r / T^4_{ }$ (for \eq\nr{k_aH_1})
can be found in ref.~\cite{eos15}.\footnote{%
 The dimensionless prefactor in \eq\nr{k_aH_1}
 can be written as 
 \be
  \frac{2\pi}{ \mbox{s}\, T^{ }_0}
  \sqrt{\frac{3}{8\pi}}
  =
  \sqrt{\frac{3\pi}{2}}
  \frac{\hbar/(\mbox{eV\hspace*{0.2mm}s})}
       {(T^{ }_0/\mbox{K})
        (k^{ }_\rmiii{B}\mbox{K}/\mbox{eV})} 
  \;, \la{k_aH_2}
 \ee
 where
 $
  \hbar = 6.582119569\times 10^{-16}_{ }\, \mbox{eV\hspace*{0.2mm}s}
 $, 
 $
   T^{ }_0 \approx 2.7255 \mbox{\hspace*{0.4mm}K}
 $,
 and
 $
  k^{ }_\rmiii{B}\hspace*{-0.1mm}K
  = 8.617333262 \times 10^{-5}_{ }
  \,\mbox{eV}
 $.}

We recall in passing that the number of $e$-folds
in the postinflationary epoch reads  
\be
 \ln \biggl( \frac{a^{ }_0}{a^{ }_e} \biggr)
 \; = \; 
 \ln\biggl(
    \frac{ T^{ }_e }{ \mbox{GeV} } 
    \frac{10^9_{ }}{
                      (T^{ }_0 / \mbox{K})
                      (k^{ }_\rmiii{B}\mbox{K}/\mbox{eV}) 
                   }
    \biggr) 
 + 
 \fr13 
 \ln\biggl(
     \frac{{s}^{ }_e/T_e^3}{{s}^{ }_0 / T_0^3} 
    \biggr) 
 \; \approx \; 
 53.4
 \;. \la{efolds_thermal}
\ee
The end of inflation,
$T^{ }_e$, is chosen at a later moment than when 
the maximal temperature, $T^{ }_\rmi{max}$, had been reached. 
For a physical result, 
we should add the $e$-folds between 
$T^{ }_e$ and $T^{ }_\rmi{max}$, 
which for the benchmark shown in 
\fig\ref{fig:benchmark} gives $\Delta N \simeq 4$.
In total the thermal epoch therefore amounts to 
$
 N^{ }_\rmi{thermal} \; \simeq \; 57
$
$e$-folds. To compensate for this, i.e.\ 
to guarantee that the modes which have 
re-entered the horizon just recently, were causally connected
before inflation, it is customary to require that the inflationary
period extended for at least 
$\sim 60$ $e$-folds before reaching $T^{ }_\rmi{max}$, 
and we adopt this convention in the following.\footnote{%
 In our benchmark, the CMB frequencies 
 $f^{ }_0 = ( 10^{-19}_{ } - 10^{-16}_{ } )$~Hz, 
 corresponding to the wave numbers 
 in footnote~\ref{Mpc},  
 crossed the horizon $\Delta N = 59 - 52$ $e$-folds before 
 the solution reached $T^{ }_\rmii{max}$, respectively. \la{fn:dN}
 } 

\vspace*{3mm}

Returning to \eq\nr{X_eq_1}, 
the last term plays little role 
when $k/\H \ll 1$, and the solution
stays close to the initial value $\partial^{ }_u X = 0$. 
Once a mode re-enters the horizon, so that $k / \H \gg 1$,
the last term dominates, 
and rapid oscillations set in. 
Their precise treatment requires care. 
For a fixed $f^{ }_0$/Hz, 
we first follow \eq\nr{k_aH_1} from low temperatures up to $T^{ }_{e}$, 
to see if the mode was outside the horizon in the first place.
If yes, we integrate from $u=u^{ }_e$ 
up to a point where $ k /\H = 30_{ }$, 
resulting in $30/(2\pi) \sim 5 $ oscillations.\footnote{%
 As long as the matching point satisfies
 $k/\H \ge 20$, the results are independent of the choice. 
 } 
At this point, following ref.~\cite{eos2}, we match
onto the asymptotic $u \gg 1$ solution, 
used for extrapolating to the present time. 

\begin{figure}[t]

\hspace*{-0.1cm}
\centerline{%
   \epsfysize=7.5cm\epsfbox{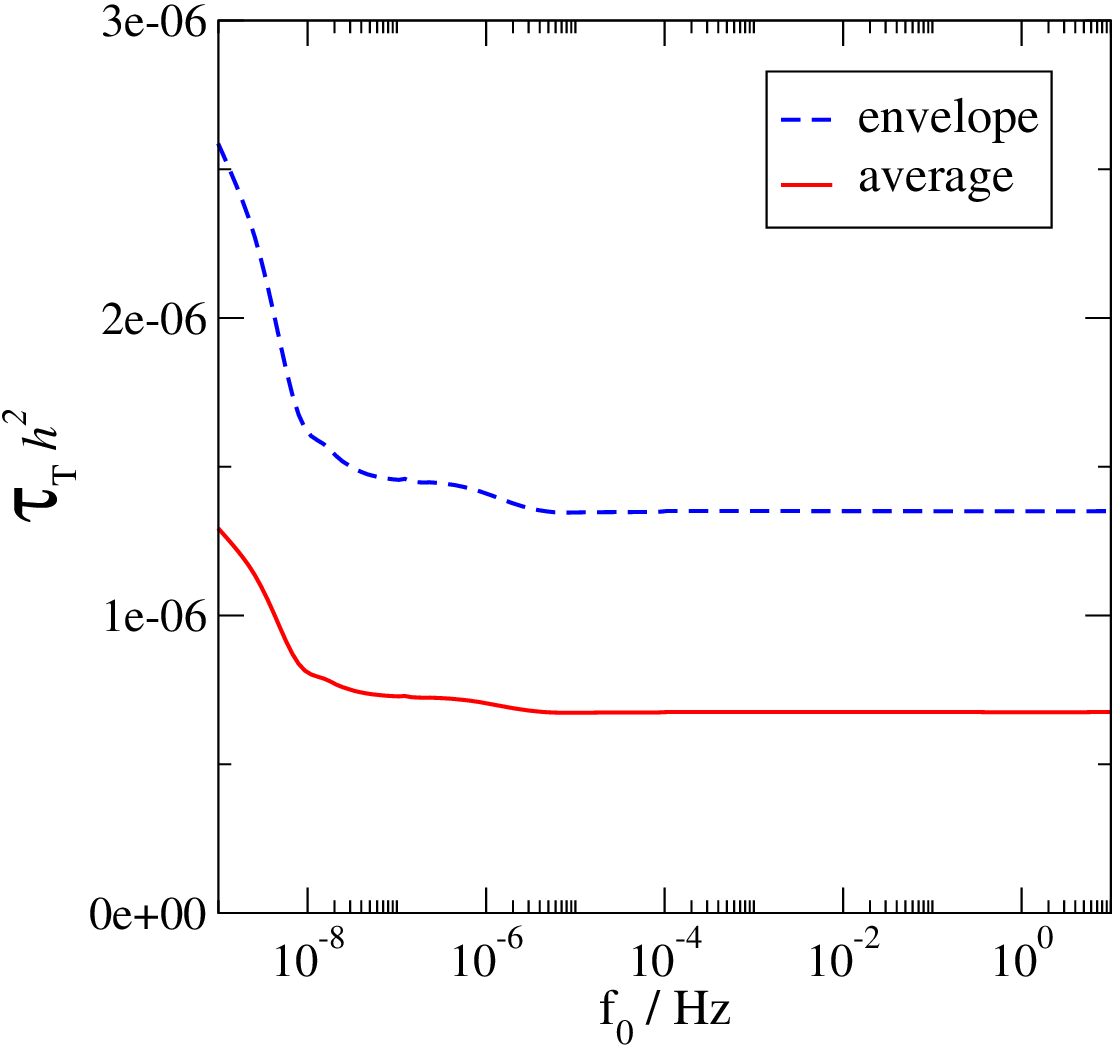}
}

\caption[a]{\small
 The transfer function from \eq\nr{calT_T_2}. 
 The dashed line shows 
 the envelope
 ($\cos^2\to 1$)
 and the solid line the average
 ($\cos^2\to \frac{1}{2}$)
 of rapid oscillations.
 The feature at $f^{ }_0 \sim 10^{-8}_{ }$~Hz originates from 
 the QCD crossover, treated according to ref.~\cite{ls};
 the smooth features  at $f^{ }_0 \sim 10^{-6}_{ }$~Hz are related to 
 the crossing of mass thresholds~\cite{ls};
 whereas the electroweak crossover,
 treated according to ref.~\cite{lm},  
 leaves behind an almost invisible shape at 
 $f^{ }_0 \sim (3-4)\times 10^{-4}_{ }$~Hz.

}

\la{fig:transferT_radiation}
\end{figure}

Let us express the asymptotic solution, 
valid for $k^2 \gg a''/a$, as
\be
 X \; {\approx} \; 
 \alpha \, 
 \frac{a^{ }_m}{a}
 \sin
 \bigl(
   u - u^{ }_m + \beta 
 \bigr) 
 \;, \la{X_asymptotics} 
\ee
where $u^{ }_m > u^{ }_e$ denotes the matching point, 
$a^{ }_m$ the scale factor at that point, 
and $\alpha$ and $\beta$ are integration constants.  
The latter can be matched onto
$X$ and $\partial^{ }_u X$ at $u^{ }_m$, 
\be
 \beta \; = \; 
 \arctan
 \biggl\{
 \biggl(
        { \frac{\H}{k} + \frac{\partial^{ }_u X}{X} } 
 \biggr)^{-1}_{ }
 \biggr\}^{ }_{u = u^{ }_m}
 \;, \quad
 \alpha = \frac{X ( u^{ }_m )}{\sin\beta} 
 \;. \la{match}
\ee
Inserting \eq\nr{X_asymptotics} and 
keeping only the dominant term
at $k \gg \H$, 
the transfer function from \eq\nr{Omega_GW} becomes 
\be
 \mathcal{T}^{ }_\rmiii{T}
 \; \approx \; 
 \frac{\alpha^2}{12}  
 \biggl( \frac{k}{a^{ }_0 H^{ }_0} \biggr)^2_{ }
 \biggl( \frac{{s}^{ }_0}{ {s}^{ }_{m}} \biggr)^{2 / 3}_{ }
 \cos^2_{ }
 \bigl( 
   u^{ }_0 - u^{ }_m + \beta 
 \bigr)
 \;. \la{calT_T_1}
\ee
Like in \eq\nr{k_aH_1}, we write $k = 2\pi a^{ }_0 f^{ }_0$, 
and insert then coefficients like in \eq\nr{k_aH_2},\footnote{%
 Here the further values
 $
  h c / ( \mbox{\hspace*{0.4mm}m} H_0^{ } )\approx 
  0.9250629\times 10^{26}_{ }  
 $
 and
 $
  c \mbox{\hspace*{0.4mm}s} / \mbox{m} = 
  2.99792458 \times 10^8_{ }
 $
 were needed. 
 } 
\be
 \mathcal{T}^{ }_\rmiii{T}\, h^2_{ }
 \; \approx \; 
 \frac{\alpha^2}{12}  
 \biggl( \frac{ f^{ }_0 }{ \mbox{Hz} } \biggr)^2_{ }
 \biggl( \,
 \underbrace{ 
               \frac{ 2\pi
                      [ h c / ( \mbox{\hspace*{0.4mm}m} H_0^{ } ) ]
                      (T^{ }_0 / \mbox{K})
                      (k^{ }_\rmiii{B}\mbox{K}/\mbox{eV}) 
                     }{ 10^9_{ } (c \mbox{\hspace*{0.4mm}s} / \mbox{m}) }
 }_{ \approx\; 4.5535 \times 10^5_{ }} 
 \, \biggr)^2_{ }
 \biggl( \frac{ \mbox{GeV} }{ T^{ }_m } \biggr)^2_{ }
 \biggl( \frac{ {s}^{ }_0 / T_0^3}
              { {s}^{ }_{m} / T_m^3} \biggr)^{2 / 3}_{ }
 \cos^2_{ }
 \bigl( 
   ... 
 \bigr)
 \;. \la{calT_T_2}
\ee
Even if $\alpha$, $T^{ }_m$ and $s^{ }_m$ depend
on the matching point, the final result is independent of it. 

Rather than showing the rapid oscillations 
from \eq\nr{calT_T_2}, we plot
their envelope
($\cos^2\to 1$)
and average
($\cos^2\to \frac{1}{2}$), 
with a result that 
can be found in \fig\ref{fig:transferT_radiation}.
According to \eq\nr{Omega_GW}, this determines the relation
between $ \P^{ }_\rmii{T} $ at $T = T^{ }_e$, 
and $ \Omega^{ }_\rmiii{GW} $ at the present time. 

%
\section{Analytical and numerical results}
\la{se:numerics}

The purpose of this section is to illustrate the theoretical
results of \se\ref{se:tensor} with semi-realistic parameter values. 
We first carry out a parameter scan of CMB constraints for the
potential of \eq\nr{V} (cf.\ \se\ref{ss:scan}); analyse 
the corresponding gravitational wave predictions in the LISA
frequency window (cf.\ \se\ref{ss:template}); 
and turn then to other potentials (cf.\ \se\ref{ss:other}).

%
\subsection{Scan of parameter space}
\la{ss:scan}

\begin{figure}[t]

\hspace*{-0.1cm}
\centerline{%
      \epsfysize=5.2cm\epsfbox{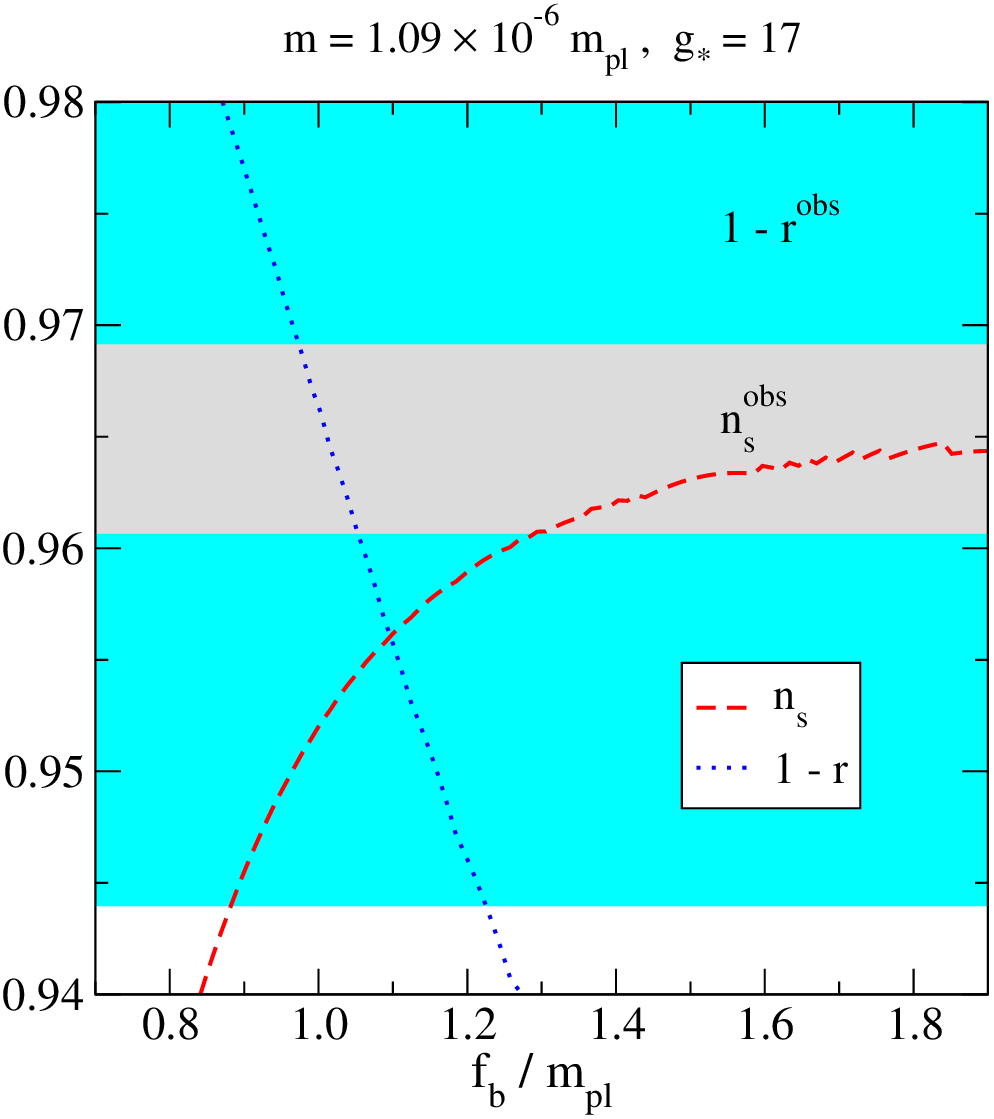}
   ~~~\epsfysize=5.2cm\epsfbox{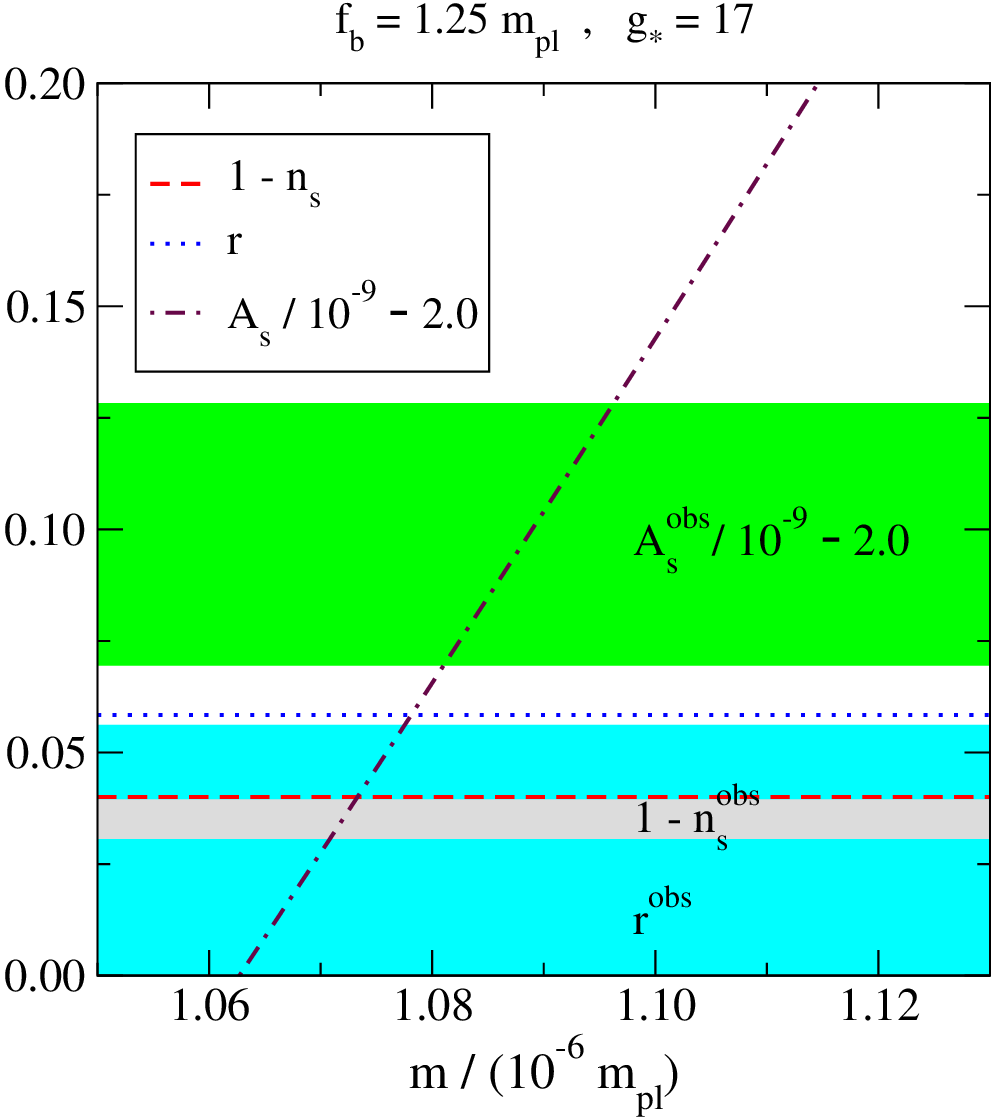}
   ~~~\epsfysize=5.2cm\epsfbox{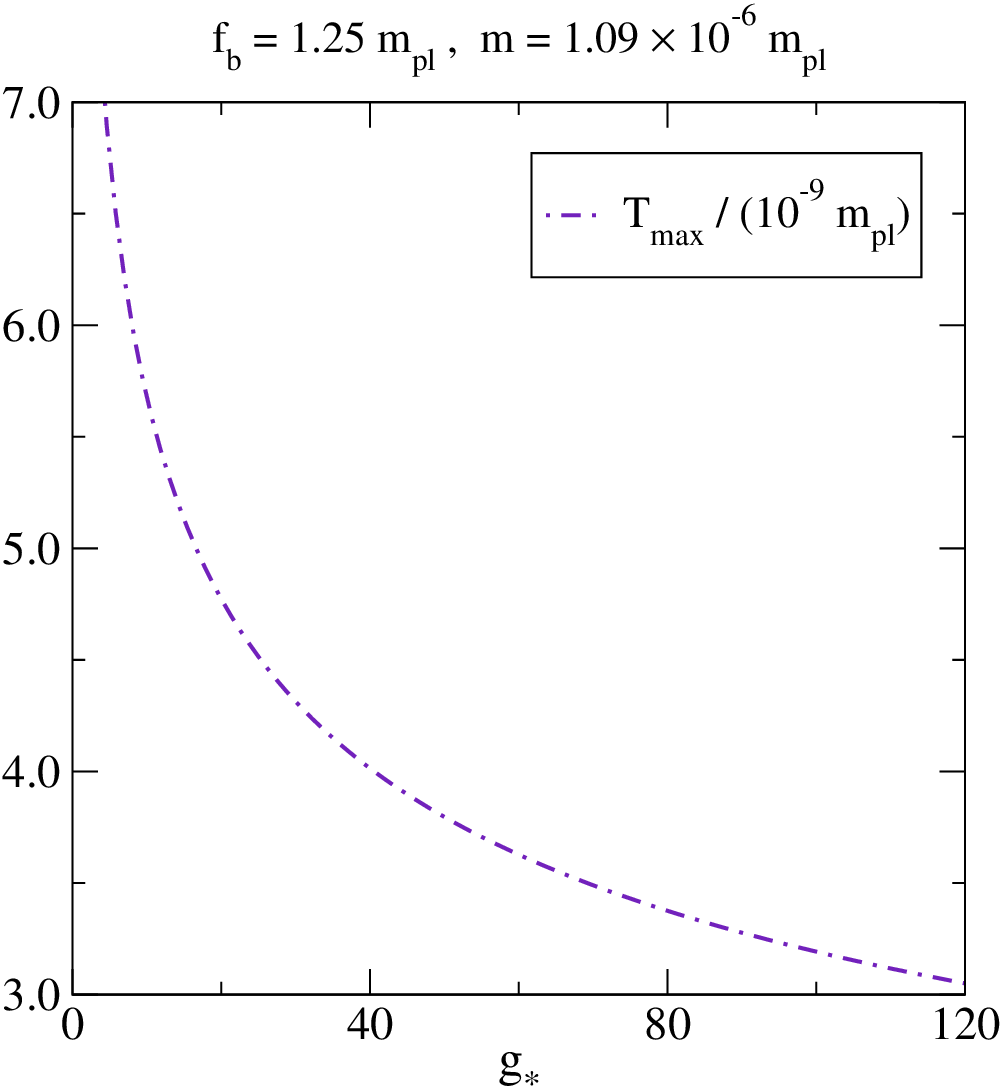}
}

\vspace*{-2mm}

\caption[a]{\small
 The observables $A^{ }_s$, $n^{ }_s$ and $r$
 (cf.\ \eqs\nr{As}--\nr{r})
 as a function 
 of 
 $ f^{ }_b / \mpl^{ } $ (left) 
 and 
 $ 
   m / ( 10^{-6}_{ }\, \mpl^{ } )  
 $ (middle) 
 for the potential in \eq\nr{V}. 
 The parameters not varied have been set to the benchmark values
 $ f^{ }_b = 1.25 \, \mpl^{ } $, 
 $ m = 1.09 \times 10^{-6}_{ }\, \mpl^{ } $ 
 and $g^{ }_* = 17$.
 The bands show the $1\sigma$ (68\% CL) contours~\cite{planck}. 
 The right panel illustrates the maximal temperature, 
 as a function of $g^{ }_*$.
}

\la{fig:scan}
\end{figure}

As the background solution (cf.\ \se\ref{se:background})  
turns out to be in a weak regime of warm inflation 
(i.e.\ with $\Upsilon \ll H$)
at the time 
when CMB perturbations are generated, 
we adopt vacuum predictions for 
scalar perturbations. 
This also concerns tensor perturbations in the frequency
domain relevant for CMB predictions, 
i.e.\ $f^{ }_0 \ll 10^{-15}_{ }$~Hz. 
The standard expressions read
\ba
 A^{ }_s & \equiv & 
 \P^{ }_\mathcal{R}(k)  
 \bigr|^{ }_{H^{ }_*}
 \; \approx \; 
 \biggl( \frac{H_*^2}{2\pi\dot{\bar{\varphi}} } \biggr)^2_{ }
 \;, \la{As} \\ 
 n^{ }_s & \equiv & 
 1 + \frac{{\rm d}\ln \P^{ }_\mathcal{R}(k) }{{\rm d}\ln k}
 \biggr|^{ }_{H^{ }_*}
 \; \approx \; 
 1 + \frac{2H^{ }_*}{H^2_{*} + \dot{H}^{ }_*}
 \biggl(
  \frac{2\dot{H}^{ }_*}{H^{ }_*} 
 - 
  \frac{\ddot{\bar{\varphi}}}{\dot{\bar{\varphi}} }
 \biggr)
 \;, \la{ns} \\
 r & \equiv & 
 \frac{ \P^{ }_\rmii{T}(k) }{ \P^{ }_\mathcal{R}(k) }
 \biggr|^{ }_{H^{ }_*}
 \; \approx \; 
 4 \pi G \biggl( 
           \frac{4\dot{\bar{\varphi}}}{H^{ }_*}
         \biggr)^2_{ }
 \;, \la{r} 
\ea
where 
$\mathcal{R}$ denotes a curvature perturbation, 
$
 k^{ }
$
is chosen as a typical CMB co-moving momentum, 
and 
$
 H^{ }_*
$
is the Hubble rate at the time when this mode first exits the horizon
($k^{ } = a^{ }_* H^{ }_*$). 
In terms of slow-roll parameters, {\it viz.}
\be
 \epsilon^{ }_\rmii{$V$}
 \; \equiv \; 
 \frac{1}{16\pi G}
 \biggl( \frac{V^{ }_\varphi}{V} \biggr)^2_{ }
 \;, \quad
 \eta^{ }_\rmii{$V$}
 \; \equiv \; 
 \frac{1}{8\pi G} \frac{V^{ }_{\varphi\varphi}}{V}
 \;, \la{slowroll}
\ee
\eqs\nr{As}--\nr{r} can be approximated as 
\ba
 A^{ }_s & \approx & 
 \frac{128\pi G^3}{3} \frac{V^3}{V_\varphi^2} 
 \; = \; \frac{8 G^2 V}{3\epsilon^{ }_\rmii{$V$}}
 \;, \la{As_slowroll} \\ 
 n^{ }_s & \approx & 
 1 + 
 \frac{\dot{\bar{\varphi}}}{H^{ }_*}
 \frac{\epsilon^{ }_\rmii{$V$}}{V}
 \,\partial^{ }_\varphi
 \biggl(
   \frac{V}{\epsilon^{ }_\rmii{$V$}}
 \biggr)
 \; \approx \;
 1 - 6 \epsilon^{ }_\rmii{$V$} + 2 \eta^{ }_\rmii{$V$}  
 \;, \la{ns_slowroll} \\ 
 r & \approx & 
 64\pi G 
 \biggl( \frac{V^{ }_\varphi}{8\pi G V} \biggr)^2_{ }
 \; = \; 
 16 \epsilon^{ }_\rmii{$V$} 
 \;. \la{r_slowroll} 
\ea
By convention we evaluate these $N=60$ $e$-folds before the solution reaches
the maximal temperature (cf.\ footnote~\ref{fn:dN}), whose value is plotted in
\fig\ref{fig:scan}(right). 

\begin{figure}[t]

\vspace*{3mm}

\hspace*{-0.1cm}
\centerline{%
   \epsfysize=3.75cm\epsfbox{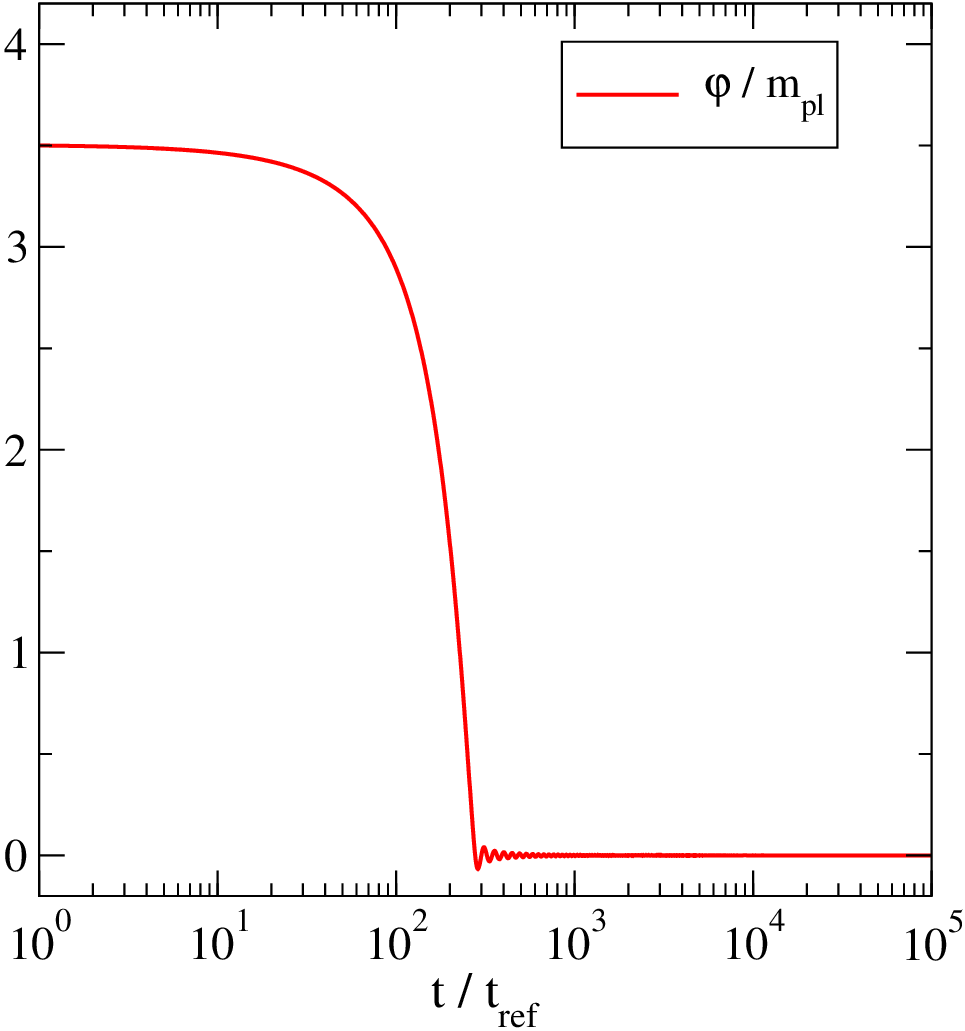}
   ~~~\epsfysize=3.75cm\epsfbox{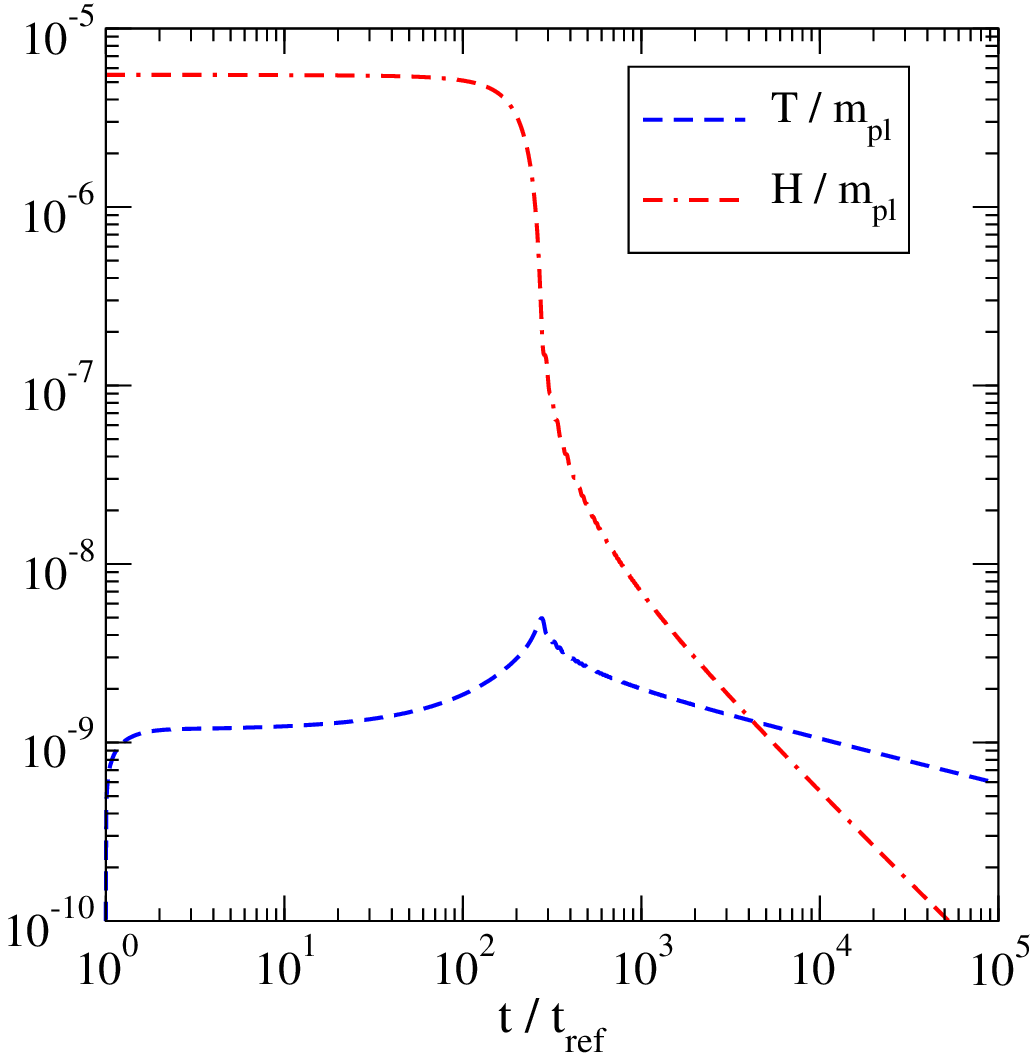}
   ~~~\epsfysize=3.75cm\epsfbox{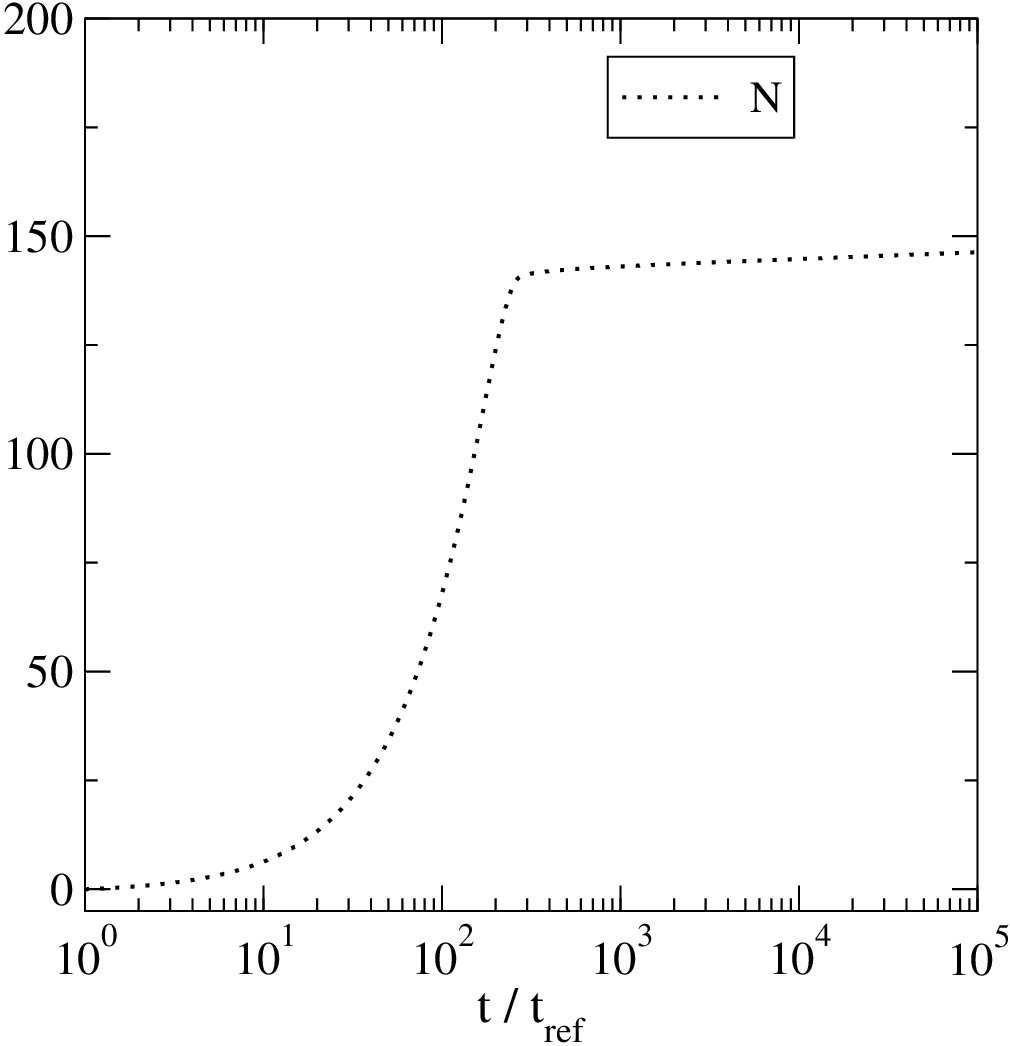}
   ~~~\epsfysize=3.75cm\epsfbox{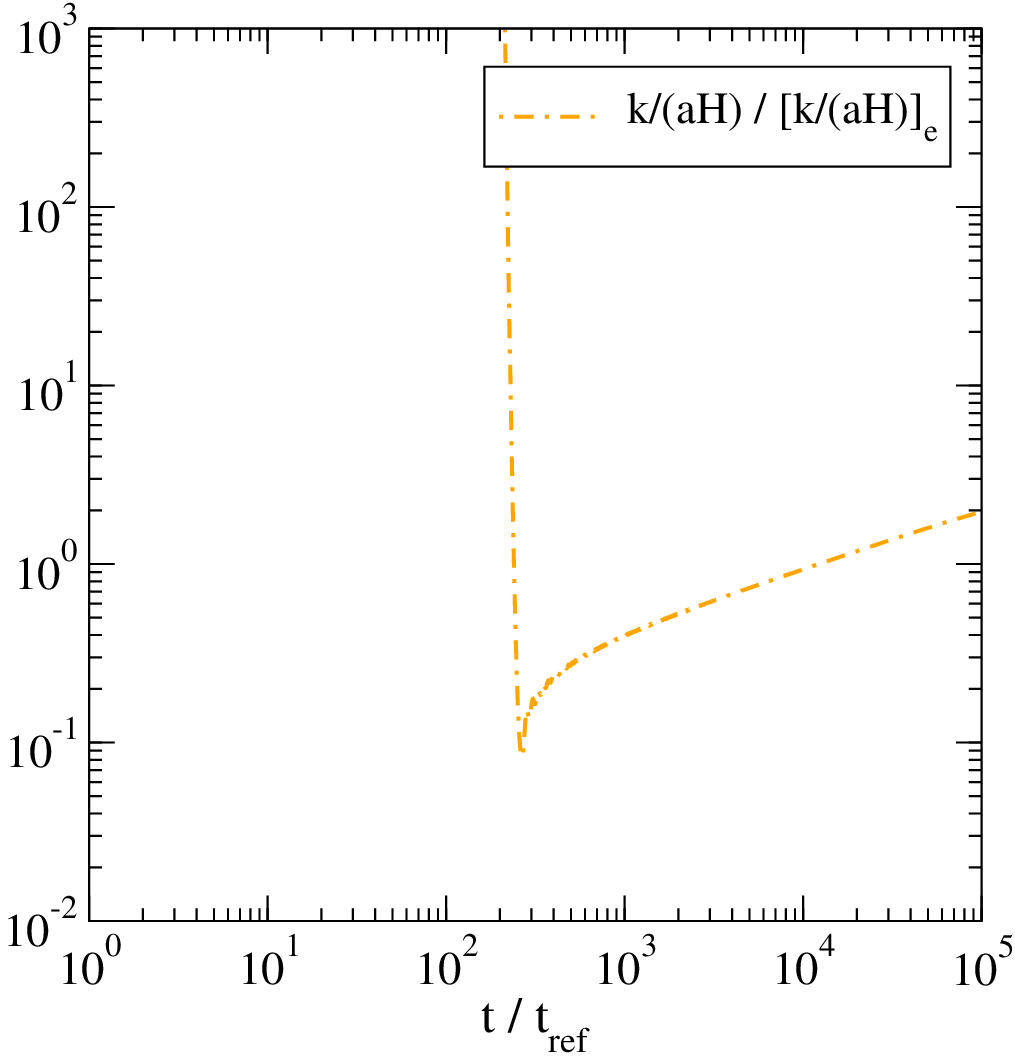}
}

\vspace*{-2mm}

\caption[a]{\small
 A benchmark solution extracted from the scans
 of \fig\ref{fig:scan}, for 
 $ f^{ }_b = 1.25 \, \mpl^{ } $, 
 $ m = 1.09 \times 10^{-6}_{ }\, \mpl^{ } $ 
 and $g^{ }_* = 17$, which lies within 
 $2\sigma$ of the observational constraints~\cite{planck}.
 The initial values have been set to 
 $ \varphi(t^{ }_\rmii{ref}) = 3.5 \,\mpl^{ } $ and 
 $ T(t^{ }_\rmii{ref}) = 10^{-10}_{ }\, \mpl^{ } $, 
 where
 $
  \scriptstyle
  t^{-1}_\rmii{ref} 
  \; \equiv \; 
  \sqrt{\frac{4\pi}{3}} 
  \frac{m \varphi(t^{ }_\rmiii{ref})}{ m^{ }_\rmiii{pl} }
 $. 
 In the last panel, 
 $ 
  k/(aH)
 $ 
 from \eq\nr{eom_koaH}
 has been normalized
 to an ``endpoint of inflation'', 
 $t^{ }_e \approx 1.2\times 10^{4}_{ } t^{ }_\rmi{ref}$ 
 (cf.\ \eq\nr{def_z}); 
 for $t > t^{ }_e$ we use
 the Standard Model radiation solution for $k/(aH)$,
 given in \eq\nr{k_aH_1}.
 We remark that $T(t^{ }_\rmii{ref})$ is quite large, 
 chosen in order not to miss a possible strong regime
 of warm inflation, however the solution is an attractor, so that
 the initial value does not affect the later behaviour.   
}

\la{fig:benchmark}
\end{figure}

We note that if we employ a potential with the shape
of \eq\nr{V}, then the mass squared drops out in the slow-roll
parameters $\epsilon^{ }_\rmii{$V$}$ and $\eta^{ }_\rmii{$V$}$
from \eq\nr{slowroll}.
Once we fix the initial field value, 
$\bar\varphi(t^{ }_\rmi{ref})$, 
close enough to the maximum of the potential, so that 
sufficiently many $e$-folds take place before reheating,
then $n^{ }_s$ and $r$ depend only on $f^{ }_b$.
This dependence is shown in \fig\ref{fig:scan}(left), 
and permits for us to fix $f^{ }_b/\mpl^{ }$.\footnote{%
 In contrast to ref.~\cite{warm_axion3}, we always set $f^{ }_a = f^{ }_b$, 
 whereby thermal effects are quite different.
 } 
Subsequently, 
$m / \mpl^{ }$ can be fixed by considering~$A^{ }_s$, as 
can be seen from \fig\ref{fig:scan}(middle).  
Two issues should be kept in mind, however: $A^{ }_s$
depends strongly on the number of $e$-folds considered; 
and radiative corrections to the scalar mass, 
proportional to couplings of the scalar and to powers of~$H^2$, 
could be substantial in de Sitter spacetime
(cf.\ the discussion in the second paragraph
of \se\ref{se:background}). 
Therefore the value of~$m/\mpl^{ }$ 
is to be considered as a qualitative indication only. 

A benchmark solution for parameters extracted from the scans
is illustrated in \fig\ref{fig:benchmark}.

%
\subsection{Shape of the gravitational wave background}
\la{ss:template}

\begin{figure}[t]

\hspace*{-0.1cm}
\centerline{%
   \epsfysize=7.5cm\epsfbox{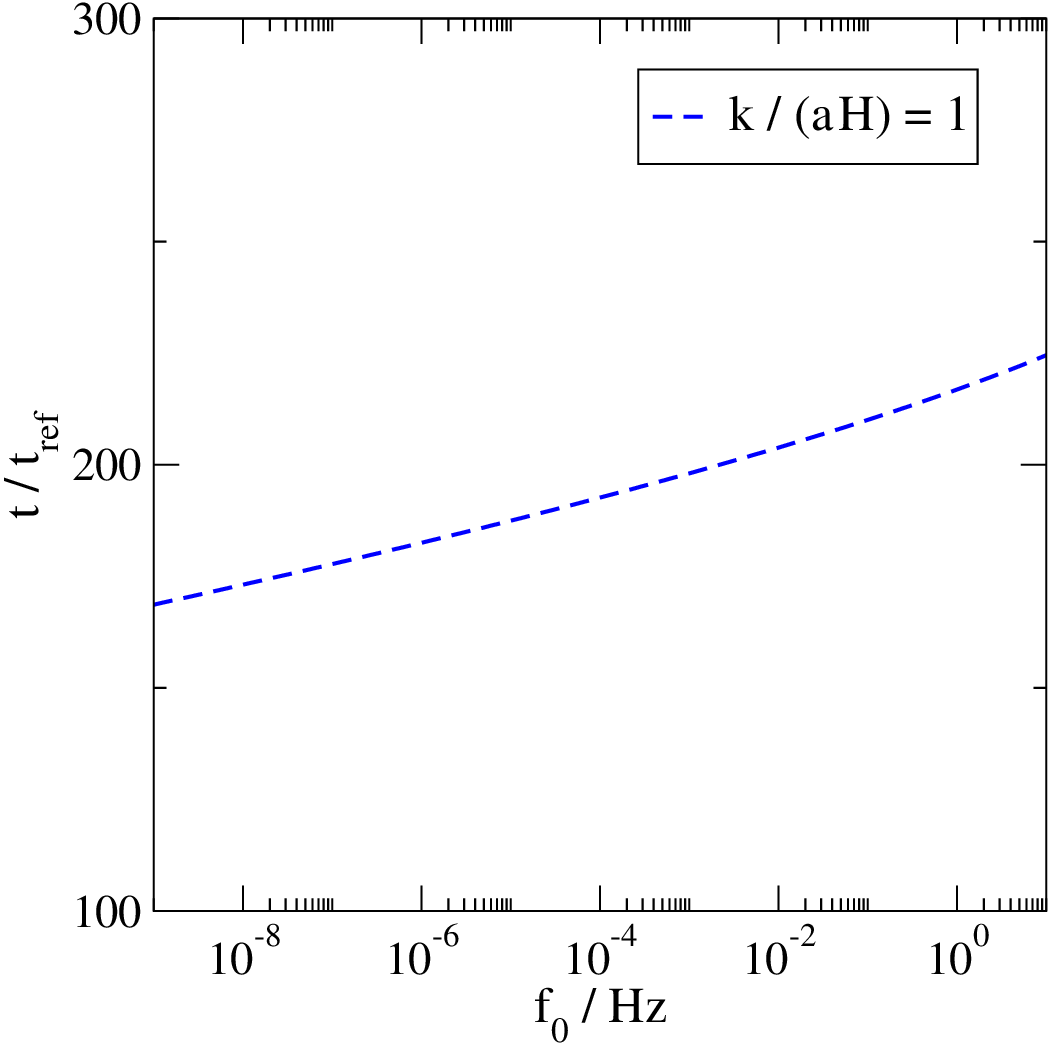}
   ~~~\epsfysize=7.5cm\epsfbox{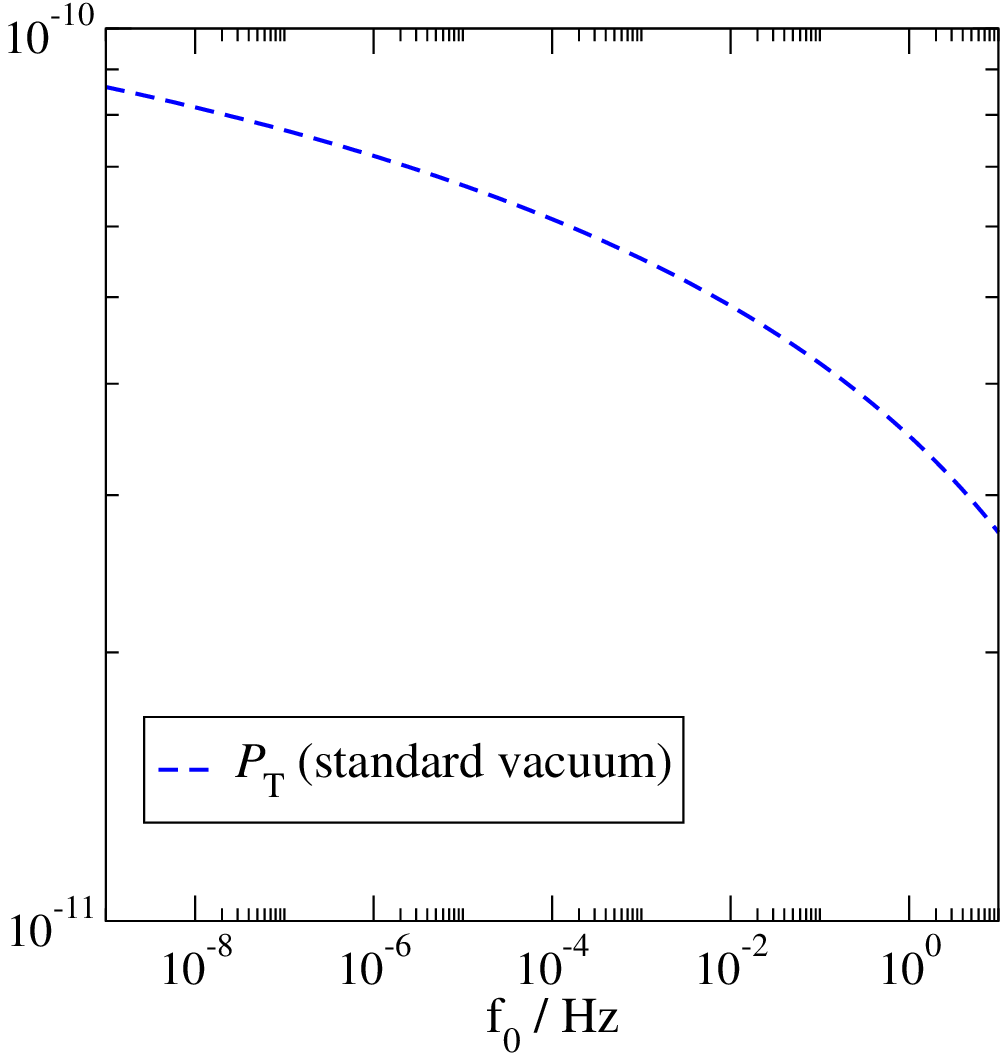}
}

\caption[a]{\small
 Left: for the solution shown in \fig\ref{fig:benchmark}, 
 the time at which a given momentum mode crosses the 
 horizon for the first time, i.e.\ $k =  a^{ } H^{ } $. 
 Right: the vacuum contribution 
 to the primordial tensor spectrum
 from \eq\nr{P_T}, originating at this moment. The non-trivial
 shape stems from the fact that large frequencies cross
 the horizon at a time when the Hubble 
 rate is already decreasing
 (cf.\ the second panel in \fig\ref{fig:benchmark}).
 For physical predictions, 
 $\P^{ }_\rmii{T}$ needs to be multiplied with 
 the transfer function from \fig\ref{fig:transferT_radiation}, 
 cf.\ \eq\nr{Omega_GW}, yielding a value 
 $\Omega^{ }_\rmiii{GW} h^2 \sim 5\times 10^{-17}_{ }$ 
 that is below LISA sensitivity~\cite{gn}.  
}

\la{fig:P_T_vac}
\end{figure}

The purpose of this section, 
which is the key part of our study,  
is to determine the gravitational wave
background that originates from a solution like in \fig\ref{fig:benchmark}.
Instead of the very small frequencies $f^{ }_0 \ll 10^{-15}_{ }$~Hz 
that affect the parameter~$r$, 
we now move to larger frequencies, in
and around the LISA window $f^{ }_0 \sim (10^{-4} - 10^{-1}_{ })$~Hz. 
To set this exercise in context, we 
note that a similar study for another model can be 
found in ref.~\cite{rt0} (an UV contribution 
viable to very large frequencies, $f^{ }_0 \gg 10^6_{ }$~Hz, 
was added in ref.~\cite{rt}).

For a general overview, let us recall 
that different contributions to 
$
 \Omega^{ }_\rmiii{GW}(f^{ }_0) 
$
lead to qualitatively different $f^{ }_0$-shapes:
\bi

\item[(i)]
The contribution from vacuum fluctuations during inflation
leads to a flat spectrum at small frequencies, \eq\nr{P_T}. 
The flat spectrum
is slightly modified by the decreasing Hubble rate
(this is illustrated in \fig\ref{fig:P_T_vac}) as well as 
the post-inflationary transfer function 
(cf.\ \fig\ref{fig:transferT_radiation}). Altogether these effects
lead to a spectral tilt $n^{ }_\rmiii{T}$, one of the 
parameters that can conceivably be constrained with 
LISA~\cite{gn}. 
Unfortunately, the overall amplitude is below 
the LISA observation threshold
(cf.\ the caption of \fig\ref{fig:P_T_vac}).  
The vacuum contribution 
is ultimately cut off at large frequencies, 
as those never crossed the horizon. 

\item[(ii)]
The contribution from thermal fluctuations, 
the second term in \eq\nr{master_P_T}, never switches off completely
but it peaks at reheating, when $T \eta$ is maximal. 
At the same time, 
this part is suppressed at small frequencies. 
A simple way to see this is to 
consider the sub-horizon production rate~\cite{gravity_qualitative}
(now written with co-moving conformal coordinates), 
\be
 \frac{{\rm d}e^{ }_\rmiii{GW}(\tau)}{{\rm d}\tau \, {\rm d}\ln k}
 \; \stackrel{k/a \lsim \alpha^2 T}{\approx} \;
 \frac{16 k^3 G T \eta }{\pi a^2}
 \;, \la{e_GW_rate}
\ee
which shows the characteristic $\sim k^3$ shape. 
The growth implies that it might be possible to satisfy
CMB constraints on the tensor-to-scalar ratio $r$ 
at very small $f^{ }_0 \ll 10^{-15}_{ }$~Hz,
yet incorporate a growing thermal part, 
perhaps approaching the LISA sensitivity window 
at larger frequencies $f^{ }_0 \sim (10^{-4} - 10^{-1}_{ })$~Hz. 

\item[(iii)]
At very large frequencies, the growing
thermal production is 
cut off at the scale $k/a\sim \pi T$, 
corresponding to the microwave frequency range $f^{ }_0 \sim 10^{11}_{ }$~Hz
today~\cite{gravity_qualitative}. Around this domain, gravitational waves
do not originate from hydrodynamic fluctuations
as considered in \se\ref{ss:thermal}, but rather from
microscopic particle collisions~\cite{gravity}.  

\ei

\begin{figure}[t]

\hspace*{-0.1cm}
\centerline{%
   \epsfysize=5.0cm\epsfbox{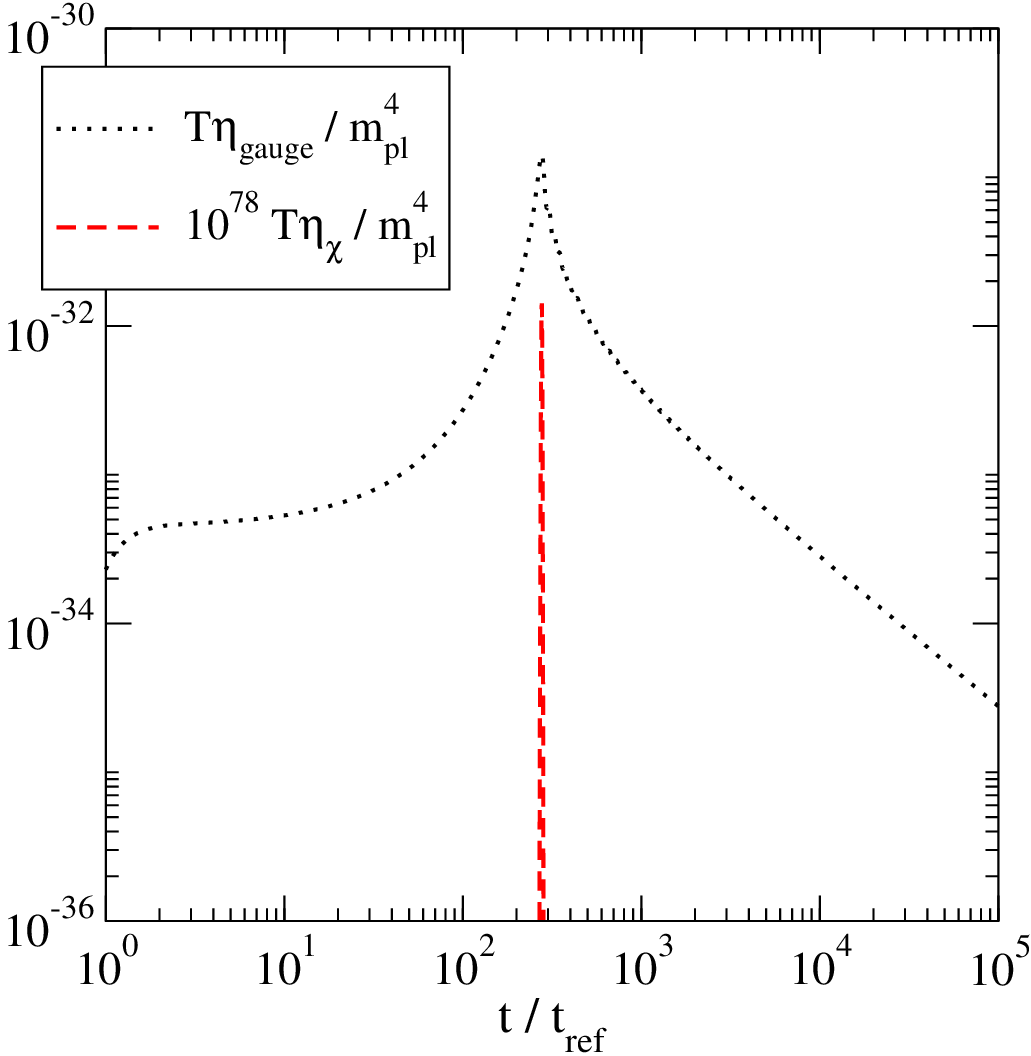}
   ~~~\epsfysize=5.0cm\epsfbox{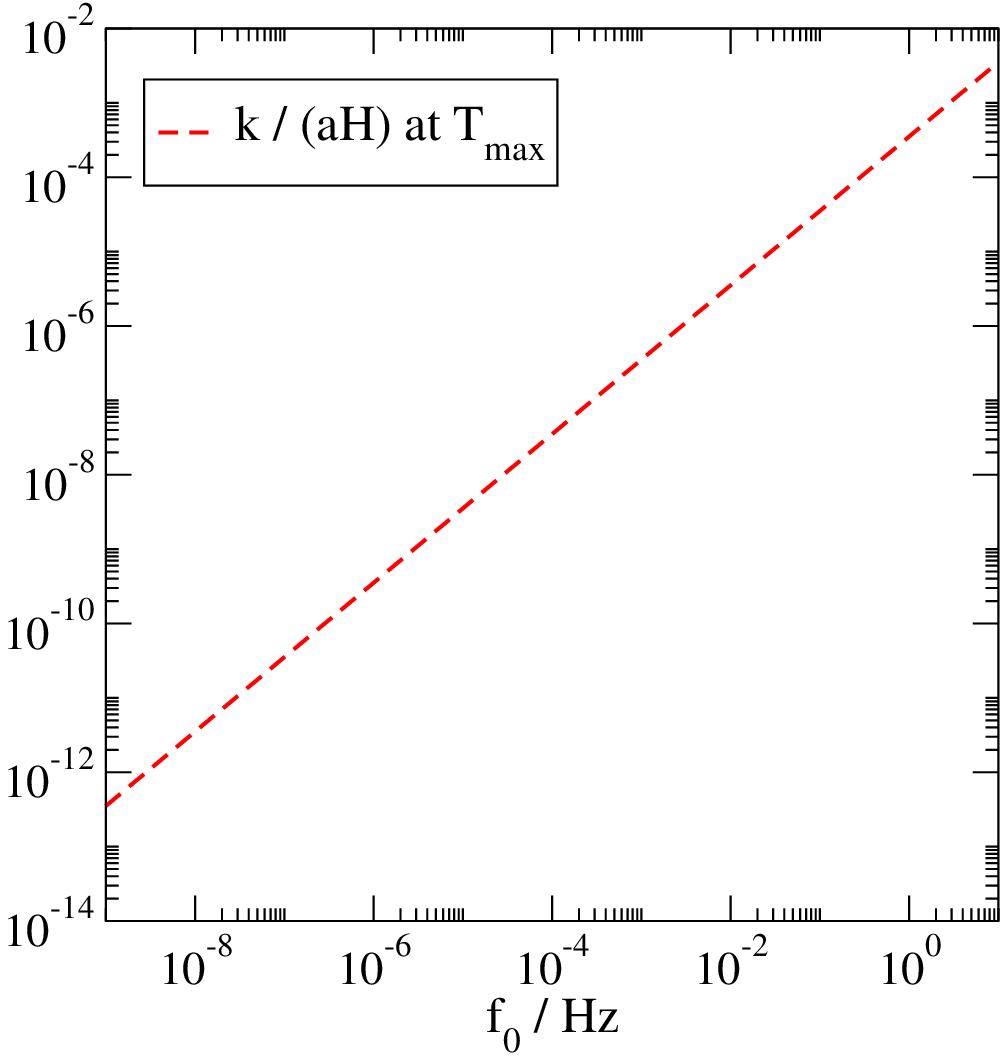}
   ~~~\epsfysize=5.0cm\epsfbox{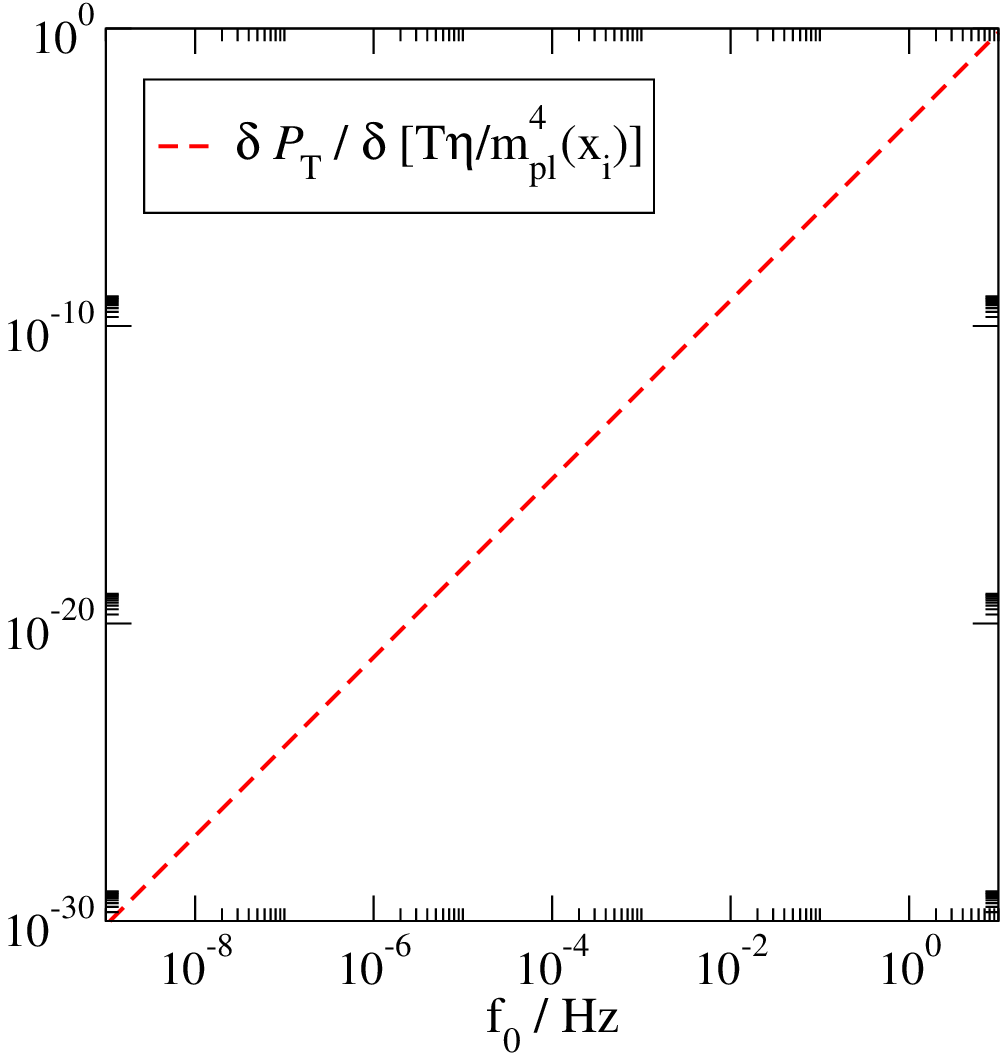}
}

\caption[a]{\small
 The ingredients influencing the thermal contribution to 
 \eq\nr{master_P_T} for the benchmark solution 
 in \fig\ref{fig:benchmark}.
 Left: 
 source terms for thermal fluctuations,
 namely temperature times the shear viscosity
 normalized to $\mpl^4$, from \eqs\nr{eta_varphi} and \nr{eta_gauge}. 
 Middle: 
 the co-moving physical momentum $k/a$
 compared with the Hubble rate at the time
 when $T\eta$ peaks, i.e.~$t \approx 276 t^{ }_\rmi{ref}$. 
 Right: 
 the (universal) transfer function between $T\eta$ 
 and the gravitational power spectrum at the  
 end of inflation, cf.\ \eq\nr{delta_master_P_T}, 
 with the time $t^{ }_i$ chosen to coincide 
 with the moment when $T\eta$ peaks.    
}

\la{fig:P_T_thermal}
\end{figure}

In order to determine the thermal contribution to 
$\P^{ }_\rmii{T}$ quantitatively, we return to \eq\nr{master_P_T}. 
Given that the thermal contribution originates at a late stage, 
when $T \eta $ peaks,
the Hubble rate is no longer constant
(cf.\ the second panel in \fig\ref{fig:benchmark}).  
Therefore we cannot use the de Sitter spacetime 
Green's function from \eq\nr{G_R_soln}. But an approximate
solution can still be found, making use of the fact that the 
relevant modes are well outside of the horizon when the maximal
$T\eta$ is reached 
(cf.\ the middle panel in \fig\ref{fig:P_T_thermal}).\footnote{%
 It may be wondered whether it is consistent to apply thermal 
 arguments to modes outside of the horizon. We do this with the 
 logic that a brief moment earlier the modes were still within the horizon, 
 and their fluctuation spectrum was determined by the corresponding dynamics. 
 } 

Making use of the logarithmic 
time variable $x$ introduced in \eq\nr{def_x}, 
we can write the differential 
contribution to \eq\nr{master_P_T} as 
\be
  \frac{ \delta\, \P^{ }_\rmii{T}(k) }
 {\delta\,  G^2 T\eta (x^{ }_i)} = 
  32^2_{ }  \, k^3 \, 
    G^{2}_\R(\taue^{ },\taui^{ },k) \, 
  \underbrace{ \frac{t^{ }_i}{a(t^{ }_i)} }
            _{ \partial \taui^{ } / \partial^{ } x^{ }_i }
  \;. \la{delta_master_P_T}  
\ee
All quantities appearing here
($k^3$, $G_\R^2$, $t^{ }_i/a(t^{ }_i)$) are dimensionless.
Imposing the initial conditions from \eq\nr{G_R_bc} 
and omitting $k^2$ from \eq\nr{tensor_T} as we are looking 
at modes with $k \ll aH$, 
the Green's function can be solved for, 
with the result
\be
 G^{ }_\R(\taue^{ },\taui^{ },k)
 \stackrel{ k \ll a H}{\approx}
 a^2_{ }(\taui^{ })
 \int_{\taui^{ }}^{\taue^{ }} \! \frac{{\rm d}\tau}{a^2(\tau)}
 \; = \; 
  a^2_{ }(t_i^{ })
 \int_{t_i^{ }}^{t_e^{ }} \! \frac{{\rm d} t }{a^3 (t)}
 \;. \la{G_R_appro}
\ee
This is independent of $k$, 
and therefore $G^{ }_\R$ is non-singular at small momenta.

Combining \eqs\nr{delta_master_P_T} and 
\nr{G_R_appro}, we establish the
scaling $k^3$ of the thermal contribution to $ \P^{ }_\rmii{T}(k) $ 
for super-horizon modes. Previously,  
\eq\nr{e_GW_rate} had shown this for sub-horizon modes, 
i.e.\ with $H \ll k/a \ll \pi T$. As the co-moving momentum $k$ is 
proportional to the current-day frequency $f^{ }_0$, 
the scaling is $\sim f_0^3$ in terms of the latter variable. 
The exact numerical solution is shown 
in the right-most panel in \fig\ref{fig:P_T_thermal},  
and reproduces this power-law. Afterwards, the spectral
shape is modified by the transfer function, 
cf.\ \fig\ref{fig:transferT_radiation}(right), but in the LISA frequency
window the modification of the shape is insignificant
(for the amplitude it is $\approx 10^{-6}_{ }$). 

We stress that the frequency 
shape $\sim f_0^3$ is universal, 
whereas its numerical coefficient is model dependent. 
The model dependence enters through the value of the shear viscosity. 
In our benchmark solution, where the temperature is small compared with
the inflaton mass, $T \ll m$, the shear viscosity originating
from interactions between the inflaton and the radiation
bath, \eq\nr{eta_varphi}, is exponentially small,
\be
 T\eta^{ }_{\chi} \; \stackrel{ T \ll m }{\approx} \;
 \frac{T^5}{\Upsilon} \biggl( \frac{m^{ }}{2\pi T} \biggr)^{3/2}_{ }
 \exp\Bigl( -\frac{m^{ }}{T} \Bigr)
 \;. \la{eta_varphi_nonrel}
\ee
The dominant contribution originates from the gauge sector, 
as given by \eq\nr{eta_gauge}. Both contributions 
are illustrated 
in the left panel of \fig\ref{fig:P_T_thermal}. 
Multiplying the left and right panels of \fig\ref{fig:P_T_thermal} 
together with the post-inflationary transfer function 
from \fig\ref{fig:transferT_radiation}(right), 
the contribution is much below the observable level. 

We end by noting that the insignificance of thermal contributions
for our benchmark solution 
can be inferred also from considering 
the second panel of \fig\ref{fig:benchmark}. 
It shows that $T \ll H$ during inflation and 
around the time when $T\eta$ peaks. This implies that 
thermal effects are sub-dominant
(cf.\ the discussion in the second paragraph
of \se\ref{se:background}). In contrast, scenarios
leading to $\alpha^2 T \gsim H$,  
which could be self-consistently treated 
with our formalism~\cite{warm}, 
might lead to a more substantial signal.

%
\subsection{Other forms of the inflaton potential}
\la{ss:other}

As explained at the end of the previous section, our benchmark
potential, \eq\nr{V}, only leads to small thermal effects for the
parameter values that are close to being
phenomenologically acceptable. We would 
now like to probe small modifications to the shape of the potential. 
Inspired by ref.~\cite{hilltop}, we do this in the minimal manner
of a small 
contribution from higher harmonics. With the constraints
$V(j\, 2\pi f^{ }_b) = V^{ }_{\varphi}(j\, 2\pi f^{ }_b) = 0$ 
and $V^{ }_{\varphi\varphi}(j\, 2\pi f^{ }_b) = m^2$, 
where $j\in\mathbbm{Z}$,  
such a change can be parametrized as 
\be
 V(\bar\varphi)
 \;\simeq\;
 m^2 f_b^2\,
 \biggl\{\,
  (1 - c)\, 
  \biggl[ 1 - \cos\biggl( \frac{\bar\varphi}{f^{ }_b} \biggr) \biggr]
  + 
  \frac{c}{n^2}\, 
  \biggl[ 1 - \cos\biggl( \frac{n\bar\varphi}{f^{ }_b} \biggr) \biggr]
 \,\biggr\}
 \;. \la{Vharm}
\ee
The basic characteristics of such potentials, 
for $n=2,3$ and allowing for both signs of $c$, 
are listed in table~\ref{table:Vharm}.
We have again treated thermal effects both by a direct change
$
 m^2 \to m_\T^2 
$, 
and by the minimal recipe of \eq\nr{V_alt}, finding 
effectively the same results. 

\begin{table}[t]

\hspace*{-0.1cm}
\begin{minipage}[c]{15.2cm}
\small{
\begin{center}
\vspace*{-4mm}
\begin{tabular}{ccccccc} 
 $ n $ & 
 $ \partial^{ }_c \{ V(\pi f^{ }_b) \}^{ }_{c=0} $ &
 $ \partial^{ }_c \{ |V^{ }_{\varphi\varphi}(\pi f^{ }_b)| \}^{ }_{c=0} $ &
 $ \partial^{ }_c \{ 1 - n^{ }_s \}^{ }_{c=0} $ &
 $ \partial^{ }_c \{ r \}^{ }_{c=0} $ &
 $ \partial^{ }_c \{ T^{ }_\rmii{max} / \mpl \}^{ }_{c=0} $ 
 \\[2mm]
 \hline
 \\[-4mm] 
 2 & $<0$ & $<0$ & $-0.011$ & $-0.035$ & $-1.6 \times 10^{-11}_{ }$    
  \\
 3 & $<0$ & $=0$ & $-0.021$ & $+0.041$ & $-3.9 \times 10^{-11}_{ }$   
  \\
 \hline 
\end{tabular} 
\end{center}
}
\end{minipage}

\vspace*{3mm}

\caption[a]{\small
     Changes resulting from the potential in \eq\nr{Vharm}, 
     compared with the case $n=1$ from \eq\nr{V}.
     The derivatives have been evaluated
     at our benchmark point 
     $ f^{ }_b = 1.25 \, \mpl^{ } $, 
     $ m = 1.09 \times 10^{-6}_{ }\, \mpl^{ } $ 
     and $g^{ }_* = 17$, keeping the number of $e$-folds 
     before $T^{ }_\rmii{max}$ fixed at $\Delta N = 60$. 
     The case $n=2$, $c>0$, which flattens the potential 
     at top, 
     brings down both $1-n^{ }_s$ and $r$,
     which is preferable for phenomenology
     (cf.\ \fig\ref{fig:scan}(middle)). 
     However, $T^{ }_\rmii{max}$ hardly changes,
     and then in the negative direction.  
}

\la{table:Vharm}
\end{table}

The upshot from table~\ref{table:Vharm} is that our
baseline results, as displayed in \fig\ref{fig:scan}, 
can be brought into $\le 1\sigma$ agreement with observational 
constraints, through a minor $n=2$ adjustment of the shape of the potential. 
However, this does not bring about larger thermal effects; 
for this, a more substantial modification of the shape
of the potential is needed. Alternatively, a larger 
$T^{ }_\rmi{max}$ could be obtained with a smaller $g^{ }_*$
(cf.\ \fig\ref{fig:scan}(right)), or by considering
a plasma in a confined phase
(cf.\ discussion between \eqs\nr{eom_friedmann} and \nr{eom_N}).

%
\section{Conclusions}
\la{se:concl}

Non-Abelian gauge fields are believed to thermalize rapidly, 
rendering warm inflation scenarios generically plausible~\cite{fixed_pt}. 
In view of this fact,
we have considered the contribution of thermal fluctuations 
to the gravitational wave background that originates
during inflation, concentrating specifically on 
the LISA frequency window. 
We have derived an interpolating formula, \eq\nr{master_P_T},
which includes both vacuum and thermal contributions, 
and is not restricted to de Sitter spacetime, which ceases to be 
a good approximation towards the end of inflation. 
In \se\ref{ss:template}, 
we have shown how the thermal part can be evaluated numerically, 
and how its main features can be understood analytically.

Our key finding is that, 
in stark contrast to the approximately 
constant vacuum contribution, 
the thermal contribution scales as
$k^3$ in terms of the co-moving momentum, or as $f_0^3$ in terms of the 
current-day gravitational wave frequency, 
cf.\ \eq\nr{delta_master_P_T}.
Since the tensor-to-scalar ratio, $r$, 
originates from very small frequencies, 
$f^{ }_0 \ll 10^{-15}_{ }$~Hz,
such a growth implies that CMB constraints 
can be respected, yet the signal
could in principle be observable in the LISA window, 
$f^{ }_0 \sim ( 10^{-4}_{ } - 10^{-1}_{ }) $~Hz. The growth
is cut off only at very large frequencies, $f^{ }_0 \sim 10^{11}_{ }$~Hz, 
where most of the gravitational energy density lies~\cite{gravity_qualitative}.

Whether the thermal part could be observable by LISA depends
on the coefficient of the~$f_0^3$ growth. 
The coefficient is proportional to the maximal 
shear viscosity of the inflaton plus radiation system. 
We have illustrated this coefficient for a particular
model, with an SU(3) plasma coupled to an axion-like inflaton
(cf.\ \fig\ref{fig:P_T_thermal}(left)). 
For the parameter values that satisfy CMB constraints, 
the effect is way too small to 
be observable. However, the situation could 
be different in other models. In particular, those
realizing the ``strong regime'' of warm inflation, 
can have a higher maximal temperature, and it is  
sustained for a longer time.

It might be worried that if we increase the coefficient of $f^3_0$, 
then we are also likely to put more energy 
into the high-$f^{ }_0$ end of the gravitational wave spectrum, 
and may ultimately face a conflict with 
$N^{ }_\rmi{eff}$, which parametrizes the overall 
energy density (cf.\ footnote~\ref{fn:Neff}). This
consideration is similar in spirit to that constraining Abelian
axion-like models~\cite{axion_gw_4}.
We note that the maximal temperature could be increased to 
$T^{ }_\rmi{max}\approx 2\times 10^{17}$~GeV without any 
concern~\cite{gravity_lo,gravity}, which is $\sim 10^7_{ }$ higher
than in our benchmark solution. Since the coefficient of $f^3_0$
scales as $T^4$ (cf.\ \eq\nr{eta_gauge}), the gravitational wave
background in the LISA window 
would be increased by a factor $\sim 10^{28}_{ }$.
In view of \fig\ref{fig:P_T_thermal}, this could bring us
close to the observation threshold around the upper end of the LISA window, 
particularly if the peak in $T\eta$ is broad. 
Were such models to be found, a quantitative study both of 
$N^{ }_\rmi{eff}$ and of the coefficient
of $f^3_0$ would be well merited. Obviously, 
like in \fig{5} of ref.~\cite{rt}, the observational prospects
are much brighter for further experimental concepts, like DECIGO.

To summarize, axion-like inflation has been used
as a motivation for the LISA physics program, asserting that 
an observable signal 
could be obtained in certain Abelian cases~\cite{axion_lisa}. 
Our study demonstrates
that this statement depends on model details, 
and is unlikely to apply to generic non-Abelian constructions. 
That said, we also find a model-independent feature in the
spectrum, 
namely a characteristic $f_0^3$ shape in the LISA frequency window. 
The coefficient of this component
would allow to measure the maximal shear
viscosity of the early universe.  
The existence of such a feature 
underlines the versatility of the physics information
that is contained in 
the primordial gravitational wave background. 

%
\section*{Acknowledgements}

We thank Saarik Kalia, Germano Nardini and Arttu Rajantie 
for helpful discussions. 
This work was partly supported by the Swiss National Science Foundation
(SNSF) under grant 200020B-188712.

%
\appendix
\renewcommand{\thesection}{\Alph{section}}
\renewcommand{\thesubsection}{\Alph{section}.\arabic{subsection}}
\renewcommand{\theequation}{\Alph{section}.\arabic{equation}}

%
\section{On dispersion theory for the medium response}

Complementing \se\ref{ss:reCR}, we recall here
the analyticity (Kramers-Kronig) 
relations governing the real and imaginary parts of the retarded 
correlator. Let us assume first that all integrals converge. 
The correlator $C^{ }_\R$ is analytic in the upper half-plane, 
from which it follows that 
\be
 C^{ }_\R(\omega + i 0^+_{ }) = 
 \int_{-\infty}^{\infty}
 \! \frac{{\rm d}\omega'}{\pi} \frac{\rho(\omega')}{\omega' - \omega - i 0^+}
 \;, \la{spectral}
\ee
where $\rho(\omega) \equiv \im C^{ }_\R(\omega + i 0^+)$ 
is the spectral function. If $\rho$ is known, $\re C^{ }_\R$
can be determined along the real axis, by taking
the real part of \eq\nr{spectral}. 
Often (particularly in vacuum computations), 
the relation is re-expressed by making use of antisymmetry, 
$
 \rho(-\omega) = - \rho(\omega)
$, 
and by introducing the variable $s \equiv \omega^2$. Then
\be
 \re C^{ }_\R(\sqrt{s}) 
 = 
 \int_0^\infty \! \frac{{\rm d}s'}{\pi} \, 
 \mathbbm{P}\, \frac{\rho(\sqrt{s'})}{s' - s}
 \;, \la{disp_0}
\ee 
where $\mathbbm{P}$ denotes the principal value. 

An immediate problem is that 
if $\rho$ does not decrease at large values of the argument, 
the integral in \eq\nr{disp_0} does not converge. In principle 
the UV side can be ameliorated by carrying out a subtraction on both sides.
For instance, if $\rho(s)$ grows less rapidly than $\sim s$,
as is typical for cross sections, then 
\be
 \frac{ \re [ C^{ }_\R(\sqrt{s}) - C^{ }_\R(0) ]  }{s}
 = 
 \int_0^\infty \! \frac{{\rm d}s'}{\pi} \, 
 \mathbbm{P}\, \frac{\rho(\sqrt{s'})}{s'(s' - s)}
 \;. \la{disp_1}
\ee

In our case, 
$\rho(\sqrt{s})$ grows as $s^2 / \ln^2(s)$~\cite{Bulk_ope}, 
so a further subtraction would be necessary. Formally we could write 
\be
 \frac{ \re [ C^{ }_\R(\sqrt{s}) - ( 1 + s\partial^{ }_s) C^{ }_\R(0) ]  }
 {s^2_{ }}
 = 
 \int_0^\infty \! \frac{{\rm d}s'}{\pi} \, 
 \mathbbm{P}\, \frac{\rho(\sqrt{s'})}{(s')^2(s' - s)}
 \;, \la{disp_2}
\ee
however now the problem has been transferred to the IR side. 
Indeed, at finite temperature, 
spectral weight decreases only slowly at small frequency, 
$\rho(\sqrt{s'}) 
{\sim} 
\sqrt{s'}$, so the integral 
in \eq\nr{disp_2} does not converge at small $s'$. 
Therefore one should rather make use of \eq\nr{disp_1}, 
and regularize the UV part.

Let us recall another aspect associated with the subtractions. 
Suppose that we add a real polynomial, say 
$a^{ }_0 + a^{ }_1 s + a^{ }_2 s^2_{ }+ ...$, to $C^{ }_\R$. 
This is analytic in the upper half-plane, and thus in principle
a viable addition. 
It gives no contribution to $\im C^{ }_\R$ along the real axis, 
i.e.\ to $\rho(\sqrt{s})$, simply because there is no imaginary part.
But it would still appear on the left-hand side
of \eqs\nr{disp_0}--\nr{disp_2}. 
This underlines the fact that dispersion relations
are meaningful only via a careful handling of subtractions. 

Now, let us illustrate the situation by assuming that the spectral function
has a Lorentzian shape at small frequencies~\cite{warm}, 
\be
 \rho(\omega) 
 \; \simeq \;
 \rho^{ }_\rmiii{IR}(\omega) 
 \; \equiv \; 
 \frac{\omega \Delta^2 \Upsilon^{ }_\rmiii{IR}}{\omega^2 + \Delta^2}
 \;. 
\ee
Then the integrals in \eqs\nr{spectral} and \nr{disp_0} converge. 
Carrying them out yields 
\be
 \re C^{ }_{\R,\rmiii{IR}}(\omega) 
 \; = \; 
 \frac{\Delta^3 \Upsilon^{ }_\rmiii{IR}}{\omega^2 + \Delta^2}
 \; = \; 
 \Delta \Upsilon^{ }_\rmiii{IR}
 \, \biggl\{ 
 1 - 
 \frac{\omega^2}{\omega^2 + \Delta^2}
 \biggr\} 
 \;. \la{reCRir}
\ee

In contrast, the subtracted relation in \eq\nr{disp_1} only 
gives information on the $\omega$-dependent part
(the second term in \eq\nr{reCRir}). 
At the same time, $C^{ }_\R(0)$ for the operator $\chi$ 
from \eq\nr{L} 
vanishes in perturbation theory. 
It is for this reason
that the constant part of \eq\nr{reCRir}
was cancelled by the addition of 
$a^{ }_0 \equiv -  \Delta \Upsilon^{ }_\rmiii{IR}$ 
in ref.~\cite{warm}. However, the consistency of this
procedure is unclear, because the growing vacuum 
part of $\rho$ was omitted. Because of these complications,  
it is more straightforward to compute 
$
 \re C^{ }_{\R} 
$ 
directly, as we did in \se\ref{ss:reCR}. 

\small{
%

}


\begin{thebibliography}{99}

\bibitem{gw_infl} 
  A.A.~Starobinsky,
  {\it Spectrum of relict gravitational radiation and
  the early state of the universe,}
  JETP Lett.\  {30} (1979) 682
  [Pisma Zh.\ Eksp.\ Teor.\ Fiz.\  {30} (1979) 719].

\bibitem{planck}
  Y.~Akrami {\it et al} [Planck Collaboration],
  {\it Planck 2018 results.\ X.\ Constraints on inflation},
  Astron.\ Astrophys.\ {641} (2020) A10
  [1807.06211].
  
\bibitem{gw_preheat1}
  S.Y.~Khlebnikov and I.I.~Tkachev,
  {\it Relic gravitational waves produced after preheating,}
  Phys.\ Rev.\ D {56} (1997) 653
  [hep-ph/9701423].

\bibitem{gw_preheat2}
  R.~Easther and E.A.~Lim,
  {\it Stochastic gravitational wave production after inflation,}
  JCAP {04} (2006) 010
  [astro-ph/0601617].

\bibitem{gw_preheat3}
  J.~Garcia-Bellido, D.G.~Figueroa and A.~Sastre,
  {\it Gravitational wave background from reheating after hybrid inflation,}
  Phys.\ Rev.\ D {77} (2008) 043517
  [0707.0839].

\bibitem{gw_preheat4}
  J.-F.~Dufaux, G.~Felder, L.~Kofman and O.~Navros,
  {\it Gravity waves from tachyonic preheating after hybrid inflation,}
  JCAP {03} (2009) 001
  [0812.2917].

\bibitem{gn}
  R.~Flauger, N.~Karnesis, G.~Nardini, M.~Pieroni,
  A.~Ricciardone and J.~Torrado,
  {\it Improved reconstruction of a stochastic gravitational
  wave background with LISA,}
  JCAP {01} (2021) 059
  [2009.11845].

\bibitem{ai}
  K.~Freese, J.A.~Frieman and A.V.~Olinto,
  {\it Natural inflation with pseudo Nambu-Goldstone bosons,}
  Phys.\ Rev.\ Lett.\  {65} (1990) 3233.

\bibitem{ai_em}
  M.M.~Anber and L.~Sorbo,
  {\it Naturally inflating on steep potentials through
  electromagnetic dissipation,}
  Phys.\ Rev.\ D {81} (2010) 043534
  [0908.4089].

\bibitem{ai_rev}
  E.~Pajer and M.~Peloso,
  {\it A review of Axion Inflation in the era of Planck,}
  Class.\ Quant.\ Grav.\ {30} (2013) 214002
  [1305.3557].

\bibitem{axion_gw_1}
  J.L.~Cook and L.~Sorbo,
  {\it Particle production during inflation and gravitational waves
  detectable by ground-based interferometers,}
  Phys.\ Rev.\ D {85} (2012) 023534; 
  {\it ibid.}\ {86} (2012) 069901 (E) 
  [1109.0022].

\bibitem{axion_gw_2}
  N.~Barnaby, E.~Pajer and M.~Peloso,
  {\it Gauge field production in axion inflation:
  Consequences for monodromy, non-Gaussianity in the CMB,
  and gravitational waves at interferometers,}
  Phys.\ Rev.\ D {85} (2012) 023525
  [1110.3327].

\bibitem{axion_gw_3}
  M.M.~Anber and L.~Sorbo,
  {\it Non-Gaussianities and chiral gravitational waves
  in natural steep inflation,}
  Phys.\ Rev.\ D {85} (2012) 123537
  [1203.5849].

\bibitem{axion_gw_4}
  P.~Adshead, J.T.~Giblin, Jr., M.~Pieroni and Z.J.~Weiner,
  {\it Constraining axion inflation with gravitational waves from preheating,}
  Phys.\ Rev.\ D {101} (2020) 083534
  [1909.12842].

\bibitem{axion_lisa}
  N.~Bartolo {\it et al}, 
  {\it Science with the space-based interferometer LISA.\ 
  IV: probing inflation with gravitational waves,}
  JCAP {12} (2016) 026
  [1610.06481].

\bibitem{axion_br_energy}
  V.~Domcke, V.~Guidetti, Y.~Welling and A.~Westphal,
  {\it Resonant backreaction in axion inflation,}
  JCAP {09} (2020) 009
  [2002.02952].

\bibitem{axion_br_simu}
  A.~Caravano, E.~Komatsu, K.D.~Lozanov and J.~Weller,
  {\it Lattice Simulations of Axion-U(1) Inflation,}
  2204.12874.

\bibitem{axion_br_fermions}
  V.~Domcke, Y.~Ema and K.~Mukaida,
  {\it Axion assisted Schwinger effect,}
  JHEP {05} (2021) 001
  [2101.05192].

\bibitem{fixed_pt}
  W.~DeRocco, P.W.~Graham and S.~Kalia,
  {\it Warming up cold inflation,}
  JCAP {11} (2021) 011
  [2107.07517].

\bibitem{linde} 
  J.~Yokoyama and A.D.~Linde,
  {\it Is warm inflation possible?,}
  Phys.\ Rev.\ D {60} (1999) 083509
  [hep-ph/9809409].

\bibitem{warm}
  M.~Laine and S.~Procacci,
  {\it Minimal warm inflation with complete medium response,}
  JCAP {06} (2021) 031
  [2102.09913].

\bibitem{gravity} 
  P.~Klose, M.~Laine and S.~Procacci,
  {\it Gravitational wave background from non-Abelian 
  reheating after axion-like inflation,}
  JCAP {05} (2022) 021
  [2201.02317].

\bibitem{setup_warm_1}
  A.~Berera,
  {\it Warm inflation,}
  Phys.\ Rev.\ Lett.\  {75} (1995) 3218
  [astro-ph/9509049].

\bibitem{setup_warm_2}
  M.~Bastero-Gil, A.~Berera, R.O.~Ramos and J.G.~Rosa,
  {\it Warm Little Inflaton,}
  Phys.\ Rev.\ Lett.\ {117} (2016) 151301
  [1604.08838].

\bibitem{warm_axion1}
  K.V.~Berghaus, P.W.~Graham and D.E.~Kaplan,
  {\it Minimal warm inflation,}
  JCAP {03} (2020) 034
  [1910.07525].

\bibitem{warm_axion2}
  S.~Das, G.~Goswami and C.~Krishnan, 
  {\it Swampland, axions and minimal warm inflation}, 
  Phys.\ Rev.\ D {101} (2020) 10
  [1911.00323]. 

\bibitem{warm_axion3}
  Y.~Reyimuaji and X.~Zhang,
  {\it Warm-assisted natural inflation,}
  JCAP {04} (2021) 077
  [2012.07329].

\bibitem{warm45}
  F.~Takahashi and W.~Yin,
  {\it Challenges for heavy QCD axion inflation,}
  JCAP {10} (2021) 057
  [2105.10493].

\bibitem{warm_axion5}
  V.~Kamali, H.~Moshafi and S.~Ebrahimi,
  {\it Minimal warm inflation and TCC,}
  2111.11436.

\bibitem{warm_axion6}
  M.~Mirbabayi and A.~Gruzinov,
  {\it Shapes of non-Gaussianity in warm inflation,}
  2205.13227.

\bibitem{hydro}
  G.~Jackson and M.~Laine,
  {\it Hydrodynamic fluctuations from a weakly coupled scalar field,}
  Eur.\ Phys.\ J.\ C {78} (2018) 304
  [1803.01871].

\bibitem{mt}
  G.D.~Moore and M.~Tassler,
  {\it The sphaleron rate in SU(N) gauge theory,}
  JHEP {02} (2011) 105
  [1011.1167].

\bibitem{clgt}
  M.~Laine, L.~Niemi, S.~Procacci and K.~Rummukainen, 
  {\it Shape of the hot topological charge density spectral function}, 
  JHEP {11} (2022) 126
  [2209.13804]. 

\bibitem{Bulk_wdep}
  M.~Laine, A.~Vuorinen and Y.~Zhu,
  {\it Next-to-leading order thermal spectral functions
  in the perturbative domain,}
  JHEP {09} (2011) 084
  [1108.1259].

\bibitem{Bulk_ope} 
  M.~Laine, M.~Veps\"al\"ainen and A.~Vuorinen,
  {\it Ultraviolet asymptotics of scalar and pseudoscalar correlators
  in hot Yang-Mills theory,}
  JHEP {10} (2010) 010
  [1008.3263].

\bibitem{stochastic}
  A.A.~Starobinsky and J.~Yokoyama,
  {\it Equilibrium state of a self-interacting scalar field
  in the de Sitter background,}
  Phys.\ Rev.\ D {50} (1994) 6357
  [astro-ph/9407016].

\bibitem{cr}
  P.~Auclair and C.~Ringeval, 
  {\it Slow-roll inflation at N3LO,}
  Phys.\ Rev.\ D {106} (2022) 063512
  [2205.12608].

\bibitem{rr}
  R.O.~Ramos and L.A.~da Silva,
  {\it Power spectrum for inflation models with quantum and thermal noises,}
  JCAP {03} (2013) 032
  [1302.3544].

\bibitem{landau9} 
  E.M.~Lifshitz and L.P.~Pitaevskii, 
  {\it Statistical Physics, Part 2}, \S88-89
  (Butterworth-Heinemann, Oxford, 1980).

\bibitem{kapusta}
  J.I.~Kapusta, B.~M\"uller and M.~Stephanov,
  {\it Relativistic theory of hydrodynamic fluctuations with
  applications to heavy-ion collisions,}
  Phys.\ Rev.\ C {85} (2012) 054906
  [1112.6405].

\bibitem{amy2}
  P.B.~Arnold, G.D.~Moore and L.G.~Yaffe,
  {\it Transport coefficients in high temperature gauge theories,
  2.\ Beyond leading log,}
  JHEP {05} (2003) 051
  [hep-ph/0302165].

\bibitem{sorbo}
  Y.~Qiu and L.~Sorbo,
  {\it The spectrum of tensor perturbations in warm inflation,}
  Phys.\ Rev.\ D {104} (2021) 083542
  [2107.09754].

\bibitem{gravity_qualitative}
  J.~Ghiglieri and M.~Laine,
  {\it Gravitational wave background from Standard Model physics:
  qualitative features,}
  JCAP {07} (2015) 022
  [1504.02569].

\bibitem{eos0}
  Y.~Watanabe and E.~Komatsu,
  {\it Improved calculation of the primordial gravitational
  wave spectrum in the standard model,}
  Phys.\ Rev.\ D {73} (2006) 123515
  [astro-ph/0604176].  

\bibitem{eos2}
  K.~Saikawa and S.~Shirai,
  {\it Primordial gravitational waves, precisely:
  the role of thermodynamics in the Standard Model,}
  JCAP {05} (2018) 035
  [1803.01038].

\bibitem{djs}
  D.J.~Schwarz,
  {\it Evolution of gravitational waves through cosmological transitions,}
  Mod.\ Phys.\ Lett.\ A {13} (1998) 2771
  [gr-qc/9709027].
  
\bibitem{sw}
  S.~Weinberg,
  {\it Damping of tensor modes in cosmology,}
  Phys.\ Rev.\ D {69} (2004) 023503
  [astro-ph/0306304].

\bibitem{eos15}
  {\tt http://www.laine.itp.unibe.ch/eos15/}
  
\bibitem{ls}
  M.~Laine and Y.~Schr\"oder,
  {\it Quark mass thresholds in QCD thermodynamics,}
  Phys.\ Rev.\ D {73} (2006) 085009
  [hep-ph/0603048].

\bibitem{lm}
  M.~Laine and M.~Meyer,
  {\it Standard Model thermodynamics across the electroweak crossover,}
  JCAP {07} (2015) 035
  [1503.04935].

\bibitem{rt0}   
  A.~Ringwald, K.~Saikawa and C.~Tamarit,
  {\it Primordial gravitational waves in a minimal model
  of particle physics and cosmology,}
  JCAP {02} (2021) 046
  [2009.02050].

\bibitem{rt}
  A.~Ringwald and C.~Tamarit,
  {\it Revealing the cosmic history with gravitational waves,}
  Phys.\ Rev.\ D {106} (2022) 063027
  [2203.00621].

\bibitem{hilltop}
  R.~Daido, F.~Takahashi and W.~Yin,
  {\it The ALP miracle: unified inflaton and dark matter,}
  JCAP {05} (2017) 044
  [1702.03284].

\bibitem{gravity_lo}
  J.~Ghiglieri, G.~Jackson, M.~Laine and Y.~Zhu,
  {\it Gravitational wave background from Standard Model physics:
  Complete leading order,}
  JHEP {07} (2020) 092
  [2004.11392].

\end{thebibliography}
\end{document}